\documentclass[%
reprint,
superscriptaddress,
showpacs,
 amsmath,amssymb,
 aps,
 pre,
]{revtex4-1}

\usepackage{graphicx}
\usepackage{dcolumn}
\usepackage{bm}

\usepackage{color}

\newcommand{\pdiff}[2]{\frac{\partial #1}{\partial #2}}

\newcommand{\new}{\nonumber\\}
\newcommand{\abs}[1]{\left|#1\right|}

\newcommand{\ox}{\overline{\bm{x}}}
\newcommand{\oy}{\overline{\bm{y}}}
\newcommand{\om}{\overline{\mu}}
\newcommand{\on}{\overline{\nu}}

\newcommand{\s}{\mathcal{S}}
\newcommand{\La}{\bm{\Lambda}}
\newcommand{\msd}{{\rm MSD}}
\newcommand{\con}{S_{\rm conf}}

\usepackage{amsmath}	
\begin{document}

\preprint{APS/123-Qed}
\title{
Mean field theory of the swap Monte Carlo algorithm
}

\author{Harukuni Ikeda}
 \email{harukuni.ikeda@lpt.ens.fr}
\affiliation{%
 Department of Physics, Nagoya University, Nagoya, Japan
}%
\affiliation{%
 IPhT, CEA/DSM-CNRS/URA 2306, CEA Saclay, F-91191 Gif-sur-Yvette Cedex, France.
}%
\affiliation{%
Laboratoire de physique th\'eorique, Ecole normale sup\'erieure, PSL
Research University, Sorbonne Universit\'es, UPMC Univ. Paris 06,
CNRS, 75005 Paris, France
}%

\author{Francesco Zamponi}
\affiliation{%
Laboratoire de physique th\'eorique, Ecole normale sup\'erieure, PSL
Research University, Sorbonne Universit\'es, UPMC Univ. Paris 06,
CNRS, 75005 Paris, France
}%
\author{Atsushi Ikeda}
\affiliation{
Graduate School of Arts and Sciences, University of Tokyo, Tokyo, Japan
}%

\date{\today}

\begin{abstract}
The swap Monte Carlo algorithm combines the translational motion with
the exchange of particle species, and is unprecedentedly efficient for
some models of glass former.  In order to clarify the physics underlying
this acceleration, we study the problem within the mean field replica
liquid theory.  We extend the Gaussian ansatz so as to take into account
the exchange of particles of different species, and we calculate
analytically the dynamical glass transition points corresponding to the
swap and standard Monte Carlo algorithms. We show that the system
evolved with the standard Monte Carlo algorithm exhibits the dynamical
transition before that of the swap Monte Carlo algorithm. We also test
the result by performing computer simulations of a binary mixture of the
Mari-Kurchan model, both with standard and swap Monte Carlo.  This
scenario provides a possible explanation for the efficiency of the swap
Monte Carlo algorithm.  Finally, we discuss how the thermodynamic theory
of the glass transition should be modified based on our results.
\end{abstract}


\maketitle


\section{Introduction}

In many materials, with decreasing the temperature or increasing the
density, the supercooled liquids dynamics shows dramatic slowing down
and eventually gets frozen without developing any crystalline
order. This is the so-called glass
transition~\cite{debene2001,cavagna2009,gotze2009,berthier2011,biroli2013}.

One of the promising theory to explain the glass transition is the
so-called random first order phase transition theory
(RFOT)~\cite{kirkpatrick1987,kirkpatrick1989,bouchaud2004,lubchenko2007,kirkpatrick2015},
which attributes the slow dynamics to the emergence of a very large
number of long-lived metastable states.  It has been shown that the RFOT
scenario holds exactly in the high dimensional
limit~\cite{parisi2010,charbonneau2017}.  In finite dimensions, there
are several systematic approximation schemes that allow one to calculate
the quantitative values of the thermodynamic
quantities~\cite{mezard1999,mezard1999f,parisi2005,parisi2010,PhysRevLett.106.135702,berthier2011mic,mangeat2016}.
The RFOT theory predicts the existence of two important transition
densities (temperatures). The first is the dynamical glass transition
point, $\varphi_d$ [see Eq.~\eqref{eq:fdef} for a precise definition of
packing fraction $\varphi$ in our model], at which exponentially many
metastable glassy states arise in the free energy landscape. At the
mean-field level, or equivalently, at the high dimension limit, the
lifetime of the metastable state is infinite and the relaxation time
diverges.  The divergent behavior of the relaxation time upon
approaching $\varphi_d$ from the liquid side is well described by the
mode coupling theory (MCT)~\cite{bengtzelius1984,gotze2009}, which was
first independently derived by kinetic theory and later integrated in
the RFOT scenario~\cite{kirkpatrick1987,maim2016}. In finite dimensions,
contrary to the high dimensional limit, the dynamical transition is
avoided and the lifetime of the metastable states is finite even above
$\varphi_d$. Above $\varphi_d$, the relaxation is controlled by the
configurational entropy, which is the logarithm of the number of
metastable states~\cite{adam1965,kirkpatrick1989,bouchaud2004}.  With
increasing the density, the configurational entropy decreases and
eventually vanishes at the thermodynamic glass transition point, the
so-called Kauzmann transition point,
$\varphi_K$~\cite{kauzmann1948}. Above $\varphi_K$, the system is
permanently trapped in the lowest free energy state, called the ideal
glass state. It is quite challenging to reach the genuine thermodynamic
glass transition point, if any, because the relaxation time of the
supercooled liquid increases very rapidly above $\varphi_d$ and the
system easily goes out of equilibrium while still being far from
$\varphi_K$. Still, several indirect evidences that support the
existence of the thermodynamic glass transition have been reported,
including a growing static correlation
length~\cite{bouchaud2004,biroli2008,berthier2012static} and an ideal
glass transition in randomly pinned
systems~\cite{cammarota2012ideal,cammarota2013random,karmakar2013random,kob2013probing,ozawa2015equilibrium}.

Even if a thermodynamic glass transition exists, it is still unclear
whether or not such a transition would be the main ingredient inducing
slow dynamics in real supercooled liquids. Indeed, a totally different
scenario to explain the slow dynamics has been proposed. The so-called
dynamical facilitation theory (DFT) claims that kinetic constraints play
an essential role in the slow dynamics of supercooled
liquids~\cite{ritort2003,chandler2010,biroli2013}. Under this
assumptions, the theory describes well the qualitative behavior of the
relaxation time in finite
dimensions~\cite{biroli2013,PhysRevX.1.021013,PhysRevLett.117.145701}.
Furthermore, on the Bethe lattice, the Fredrickson-Andersen (FA)
model~\cite{PhysRevLett.53.1244}, which is a typical model in the DFT
class, exhibits very similar behavior to the
MCT~\cite{sellitto2005,sellitto2010dynamic,ikeda2015fre,sellitto2015,de2016,ikeda2017}.
Also, in finite dimensions, a Kac version of the FA model describes well
the avoided dynamical transition~\cite{berthier2012finite}.  These
successes of the DFT suggest that dynamic rules are indeed important.

To clarify the effects of the dynamic rules on the slow dynamics of
supercooled liquids, it is helpful to observe the dynamic rule
dependence. If the slow dynamics is originated solely by a thermodynamic
glass transition, the dynamical behavior of supercooled liquids should
depend only weakly on the details of the rules governing the
dynamics~\cite{wyart2017does}.  This assumption is however inconsistent
with recent results obtained by computer simulations using the swap
Monte Carlo algorithm (swap
MC)~\cite{gazzillo1989,PhysRevE.63.045102,cavagna2012,gutierrez2015,berthier2016,PhysRevX.7.021039,berthier2017breaking}.
The swap MC combines the standard Monte Carlo algorithm with the
exchange of particles species.  The computer simulations of some
polydisperse systems demonstrate that the swap MC can equilibrate the
system about 10 orders of magnitude faster than the standard
MC~\cite{berthier2016,PhysRevX.7.021039,berthier2017breaking}.  Clearly,
the standard RFOT scenario fails to explain this result, because it
totally neglects the details of the systems dynamics. It is thus
desirable to reformulate the RFOT scenario so as to take into account
the effects of the dynamic rule.

In this work, we perform a first step in this direction, by
investigating the binary Mari-Kurchan (MK)
model~\cite{mari2011,charbonneau2014b}, which is a mean-field model
belonging to the RFOT class, with the swap and standard MC
algorithms. We separately calculate the dynamical glass transition point
with the swap MC, $\varphi_d^{\rm swap}$, and with the standard MC,
$\varphi_d^{\rm mc}$, and we show that $\varphi_d^{\rm
mc}<\varphi_d^{\rm swap}$.  We also perform computer simulations of the
binary MK model and compare with the analytical result, showing in
particular that between $\varphi_d^{\rm mc}$ and $\varphi_d^{\rm swap}$,
the swap MC is more efficient than the standard MC.  Finally, we discuss
the thermodynamic glass transition and the relaxation dynamics above the
dynamical transition point of more realistic glass forming systems.

The organization of the paper is as follows. In
Sec.~\ref{180723_18Aug17}, we roughly sketch the main idea of our
theory. In Sec.~\ref{180812_18Aug17}, we introduce the model.  In
Sec.~\ref{180053_3Aug17}, we derive the analytical expression of the
free energy.  In Sec.~\ref{145947_4Aug17}, we calculate the order
parameters and phase diagram from the free energy.  In
Sec.~\ref{183549_18Aug17}, we report the computer simulation and compare
with the theoretical results. In Sec.~\ref{180943_12Aug17}, we discuss
the configurational entropy, the thermodynamic glass transition point,
and the activated dynamics.  In Sec.~\ref{183757_18Aug17}, we summarize
the results and conclude the work.

\section{Sketch of the framework}
\label{180723_18Aug17} 

Before going into the details of the theory and the model, here we give
 a qualitative explanation of our theory.  Within the RFOT scenario, the
 slow dynamics is attributed to the emergence of long-lived glassy
 metastable states.  The dynamics within one of these states is assumed
 to be arrested on the experimental time scale. This dynamical arrest
 manifests itself in the two-time correlation functions, such as the
 mean square displacement or the intermediate scattering functions. In
 particular, if $\bm{x}_i(t)$ is the position of particle $i$ at time
 $t$, the mean square displacement (MSD) is defined as
 \begin{align}\label{eq:MSDdef}
 \msd(t) &\equiv \frac{1}{N}\sum_{i=1}^N \left\langle
 \left(\bm{x}_i(t)-\bm{x}_i(0)\right)^2 \right\rangle.
\end{align}
In the liquid phase, the long-time limit of the MSD is diffusive,
$\msd(t) \sim 2d D t$ for $t\to\infty$, where $D$ is the diffusion
constant.  In the glass phase, instead (at least in the mean field
limit), the long time limit of the MSD is a constant, expressing the
fact that particles are caged by their neighbors and cannot diffuse, so
they remain close to their initial position at all times.  One can think
of the swap algorithm as a dynamics in which a given particle does not
have a definite type and it can exchange its type with other particles
during the dynamical evolution~\cite{PhysRevX.7.021039}. The question is
therefore whether such an exchange process can facilitate the dynamical
relaxation, leading to an increased efficiency of the swap algorithm.
 
 \begin{figure}[t]
\includegraphics[width=8cm]{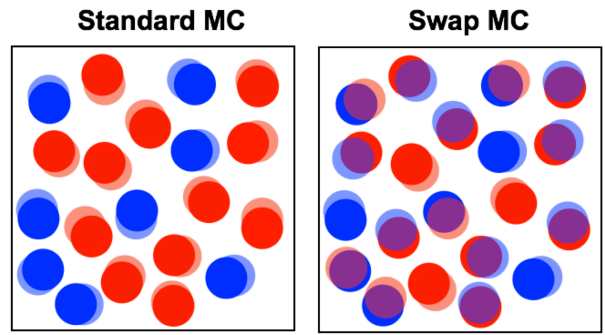}
 \caption{\small Schematic representation of the dynamics in a glass
 state, and its translation into the replica framework.  Here, the
 circles describe the position of particles, while the different colors
 describe the different species. The transparent and non-transparent
 symbols correspond to the initial configuration (encoded by replica 1)
 and the long-time configuration (encoded by replica 2),
 respectively. In the standard MC ansatz, we assume that in the glass
 phase particles keep their identity at all times (i.e. exchanges are
 forbidden), so that in different replicas, the same particle must of
 the same type. In the swap MC phase, we assume that in the glass phase
 particles can change type during time, so that in different replicas,
 the same particle can have different types.  } \label{165328_12Aug17}
\end{figure}

 If the dynamical arrest is related to the emergence of metastable
 states, it can be captured by a purely thermodynamic calculation: this
 is indeed the essence of the RFOT scenario. Such a calculation can be
 performed via the so-called replica liquid theory
 (RLT)~\cite{monasson1995}, in which long-time correlations in the
 dynamics are translated in correlations between different copies of the
 original system (replicas); see
 Refs.~\cite{mezard1999,mezard1999f,parisi2010} for details.  We thus
 introduce replicas, which can be thought as configurations of the same
 glass separated by an extremely long time evolution.  In the liquid
 phase, as the system diffuses, the different replicas are uncorrelated.
 Above the dynamical glass transition point, $\varphi_d$, many
 metastable states arise. In the metastable states, because diffusion is
 arrested, the same particles of different replicas remain close
 together, and can be thus thought as
 ``molecules''~\cite{mezard1999,mezard1999f,parisi2010}.

Our central assumption is that different kind of molecules describe different 
 dynamical rules, as illustrated in Fig.~\ref{165328_12Aug17}.
 \begin{itemize}
 \item
 In the case of the standard MC without particle swap, 
 we assume that particles of different species can not be exchanged in a
glassy metastable state, or more precisely, that the typical time scale to exchange
 particles of different species is much longer than the lifetime
 of the metastable state itself.
Hence, a particle keeps its identity at all times, and as a consequence, 
 the replica molecules must consist of
 particles of the same species, see the left cartoon of
 Fig.~\ref{165328_12Aug17}. 
  \item
 This assumption is inappropriate in the case of
 the swap MC, because particles of different species can be exchanged much
 more easily than in the standard MC. In this case, we thus
 assume that in the glass phase, particles can still change their identity over time,
and the replica molecules can thus consist of particles of different
 species, see the right cartoon of Fig.~\ref{165328_12Aug17}. This kind
 of ansatz was first proposed by Coluzzi {\it et al.}~\cite{coluzzi1999} and
 later explicitly implemented by Ikeda {\it et
 al.}~\cite{ikeda2016note}.  
\end{itemize}
The assumptions that the long-time correlations are different in the
standard and swap MC dynamics, and that they can be encoded in different
replica ansatzes, are the crucial working hypothesis behind our
analysis; we believe that it should hold at least in the mean field
limit.  In Sec.~\ref{180053_3Aug17}, we discuss the details of the two
ansatzes and calculate $\varphi_d$ of the binary Mari-Kurchan model (MK
model).  In Sec.~\ref{183549_18Aug17}, we report a comparison of the
theory with computer simulations of the same system, which provides
strong support for the validity of our hypothesis in this system.

 \section{Model}
 \label{180812_18Aug17} 
 
 In this section, we introduce the model.  We consider a system
consisting of an equal number of large and small particles.  The
particles interact with the following potential:
\begin{align}
 V_N &= \sum_{i<j} v_{\mu_i\mu_j}(\bm{x}_i-\bm{x}_j+\La_{ij}),
\end{align}
where 
\begin{align}
 v_{\mu_i\mu_j}(\bm{x}_i-\bm{x}_j+\La_{ij}) &= 
\begin{cases}
 \infty & \text{if } \abs{\bm{x}_i-\bm{x}_j+\La_{ij}}\leq \sigma_{\mu_i\mu_j}\\
 0 & \text{if } \abs{\bm{x}_i-\bm{x}_j+\La_{ij}} > \sigma_{\mu_i\mu_j}\\
\end{cases}.
\end{align}
Here, $\bm{x}_i,\bm{x}_j \in \mathbb{R}^d$ denote the particle positions and
$\mu_i,\mu_j \in \{L,\ S\}$ denote the particle species.  $\sigma_{LL}$,
$\sigma_{SS}$ are the diameters of large and small particles,
respectively. We assume that the potential is additive,
$\sigma_{LS}=\sigma_{SL}= (\sigma_{LL}+\sigma_{SS})/2$.  $\La_{ij}$ is a
quenched randomness and for each pair of $i<j$, $\La_{ij}$ is generated
independently from the probability distribution function,
\begin{align}
 P(\La_{ij}) = \frac{1}{V},\label{174243_1Aug17}
\end{align}
where $V$ is the volume of the system.  The total number of particles is
$N = N_L + N_S$, with particle concentrations $x_\mu = N_\mu/N$.  The
number density is $\rho = N/V$, and the packing fraction is given, in
the case $d=3$ which will be our focus in the following, by
\begin{equation}\label{eq:fdef}
\varphi =\frac{\pi}6 \rho [ x_L \sigma_{LL}^3 + x_S \sigma_{SS}^3 ] \ . 
\end{equation}
 We also impose $\La_{ij}=
-\La_{ji}$ so that the shifted distance between the $i$-th and $j$-th
particles is to be symmetric,
\begin{align}
\abs{\bm{x}_i-\bm{x}_j+\La_{ij}}
 = \abs{\bm{x}_j-\bm{x}_i+\La_{ji}}.
\end{align}
Because of the quenched randomness, $\La_{ij}$, particles interact with
other randomly chosen particles instead of their nearest neighbor
particles. The model is similar to models defined on random interaction
graphs, and one can obtain the analytical expression of the free energy
through mean field techniques~\cite{mari2011}.

\section{Free energy calculation}
\label{180053_3Aug17} 

In this section, we derive the analytical expression of the free energy.
In case of the swap MC, one should take into account the exchange of the
particle species as well as the translational motion of the particle
position.  The partition function can be written as
\begin{align}
 Z[\{\La_{ij}\}] &= \prod_{i=1}^N \sum_{\mu_i}\int d\bm{x}_i
 e^{-\beta \sum_{i<j}v_{\mu_i\mu_j}(\bm{x}_i-\bm{x}_j+\La_{ij})},\label{212830_1Aug17}
\end{align}
where $N$ is the number of particles and $\beta$ is the inverse
temperature.  Note that the Gibbs factor $N!$ does not appear, because
all particles are distinguishable due to the quenched
randomness~\cite{mari2011}.  Using the self-averaging property, the free
energy can be calculated as
\begin{align}
-\beta F \equiv \log Z[\{\La_{ij}\}]\approx
 \overline{\log Z[\{\La_{ij}\}]},\label{222152_1Aug17}
\end{align}
where the overline denotes averaging over $\La_{ij}$.  We analyze the
free energy using the replica method~\cite{megard1987,nishimori2001}. Because of
the quenched disorder, the treatment might look different from
the usual replica liquid theory~\cite{parisi2010}. However,
as we will see below, the two methods are identical. To perform the disordered
overage, we rewrite Eq.~(\ref{222152_1Aug17}) as
\begin{align}
-\beta F 
 &= \lim_{n\to 0}\frac{\log \overline{Z[\{\La_{ij}\}]^n}}{n}.\label{222807_1Aug17}
\end{align}
We shall use the one-step replica symmetric breaking ansatz (1RSB); we
divide $n$ replicas into $n/m$ subgroups and assume that only the $m$
replicas in the same group are
correlated~\cite{megard1987,nishimori2001}.  The 1RSB structure, coupled
to the fact that replicas in different blocks are completely
uncorrelated (this property does not hold for all models), allows
to factorize the partition function as $\overline{Z^n}=
(\overline{Z^m})^{\frac{n}{m}}$. Substituting this expression into
Eq.~(\ref{222807_1Aug17}), one obtains
\begin{align}
-\beta F &= \frac{\log \overline{Z^m}}{m},\label{012749_2Aug17}
\end{align}
where 
\begin{align}
 \overline{Z^m} &= \prod_{i<j}\int d\La_{ij}P(\La_{ij})\left(\prod_{a=1}^m\prod_{i=1}^N \sum_{\mu_i^a}\int d\bm{x}_i^a\right)\new
 &\times  \exp \left[
 -\beta \sum_{a=1}^m\sum_{i<j}v_{\mu_i^a\mu_j^a}(\bm{x}_i^a-\bm{x}_j^a+\La_{ij})
 \right].\label{145212_2Aug17}
\end{align}
Note that except for the factor $m^{-1}$, the free energy
Eq.~(\ref{012749_2Aug17}) is the same of the one considered in the standard
replica liquid theory~\cite{parisi2010}. Thus, we can use the standard
RLT of usual supercooled liquids without the quenched disorder.
The partition function
Eq.~(\ref{145212_2Aug17}) can be analyzed using the saddle point method,
see Ref.~\cite{mari2011} for the details. After some straightforward
calculations, we obtain
\begin{align}
 \s_m \equiv \frac{\log \overline{Z^m}}{N} &= \log N -\frac{1}{N}\sum_{\om}\int d\ox \rho_{\om}(\ox)\log\rho_{\om}(\ox) \new
 &+ \frac{1}{2N}\sum_{\om,\on}\int d\ox d\oy \rho_{\om}(\ox)\rho_{\on}(\oy)f_{\om,\on}(\ox-\oy),\label{145416_2Aug17}
\end{align}
where we have used the shorthand notations $\ox=\{\bm{x}^1,\bm{x}^2,\cdots,\bm{x}^m\}$ and $\om=\{\mu^1,\mu^2,\cdots,\mu^m\}$.
We have also introduced the density distribution function,
\begin{align}
 \rho_{\om}(\ox) &= \sum_{i=1}^N\left\langle
 \prod_{a=1}^m \delta(\bm{x}^a-\bm{x}_i^a)\delta(\mu^a,\mu_i^a)
 \right\rangle,
\end{align}
and the replicated Mayer function,
\begin{align}
 f_{\om\on}(\ox-\oy) &= \prod_{a=1}^m e^{-\beta v_{\mu^a\nu^a}(\bm{x}^a-\bm{y}^a)}-1.
\end{align}
The full optimization of the free energy Eq.~(\ref{145416_2Aug17}) for a
completely general form of $\rho$ is a very difficult task. In order to
simplify the calculations, below we approximate $\rho$ by assuming that
it has a Gaussian form.

\subsection{Ansatz for the swap Monte Carlo algorithm}
\label{165607_3Aug17} 

\begin{figure}[t]
\includegraphics[width=8cm]{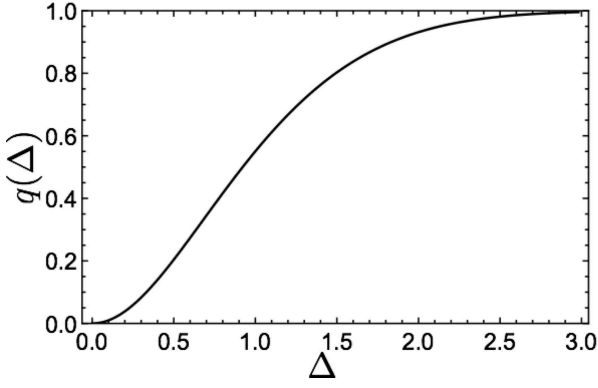}
 \caption{\small $\Delta$ dependence of $q(\Delta)$ }
 \label{qvsd}
\end{figure}

Here, we construct an ansatz for the swap MC. We
simply assume that the distribution function can be factorized as
\begin{align}
 \rho_{\om}(\ox)=\rho(\ox)g(\om).\label{193843_11Aug17}
\end{align}
In principle, one can avoid this assumption and consider a more general
ansatz, but the calculation becomes more involved as shown in
Appendix. For the distribution function of the positions, we assume a
Gaussian form~\cite{mezard1999,parisi2010}
\begin{align}
  \rho(\ox) &= \rho \int d\bm{X}\prod_{a=1}^m \gamma_{A}(\bm{x}^a-\bm{X}),\label{181209_3Aug17}
\end{align}
where $\gamma_A(\bm{x}) = e^{-\abs{\bm{x}}^2/2A}/(2\pi A)^{d/2}$. This
is the same of that used for the one-component MK
model~\cite{mari2011,parisi2010}, and we stress that this ansatz has no
particular physical meaning, it is chosen only to make the calculation
simpler.  The cage size $A$ corresponds to the order parameter of the
particle position. $A\to\infty$ corresponds to the liquid state, while a
finite value of $A$ corresponds to the glass state.  For $g(\om)$, we
assume the same form of the distribution function of the mean-field spin
glasses~\cite{megard1987,yoshino}:
\begin{align}
 g(\om) &= C_m^{-1}e^{H\sum_a \sigma(\mu^a)
 + \frac{\Delta^2}{2}\sum_{ab}\sigma(\mu^a)\sigma(\mu^b)}\new
 &= C_m^{-1}\int_{-\infty}^{\infty} Du \prod_{a=1}^me^{(H+u)\sigma(\mu^a)},\label{201047_2Aug17} 
\end{align}
where $\sigma(L)=+1$, $\sigma(S)=-1$ (i.e. large particles are
associated to up spins, small particles to down spins), and $Du =
du\times e^{-u^2/2\Delta^2}/\sqrt{2\pi \Delta^2}$.  $C_m$ is determined
from the normalization condition, $\sum_{\om}g(\om)=1$.  $H$ fixes the
numbers of large and small particles by
\begin{align}
 N_\mu &= N\sum_{\om}g(\om)\delta(\mu^a,\mu), 
\end{align}
where $N_L$ ($N_S$) is the number of large (small) particles.  In
particular for the equimolar system, $N_L=N_S$, which we shall
investigate below, one can show that $H=0$.  The correlation function of
the particles species can be calculated as a function of $m$ and
$\Delta$:
\begin{align}
 q_m(\Delta) &\equiv \left\langle \sigma(\mu^1)\sigma(\mu^2)\right\rangle
 = \frac{\int Du \tanh(u)^2\cosh(u)^m}{\int Du \cosh(u)^m} \ . \label{152834_12Aug17}
\end{align}
In equilibrium, the order parameter of the glass transition is calculated by setting
$m=1$~\cite{parisi2010}.  The function $q(\Delta)=q_1(\Delta)$ monotonically
increases with $\Delta$ from zero to unity as shown in Fig.~\ref{qvsd}.
$q$ (or $\Delta$) plays the role of the order parameter of the particle
species. $q=0$ (or $\Delta=0$) corresponds to molecules made of
completely uncorrelated particle types, and a finite value of $q$ (or
$\Delta$) corresponds to molecules made of predominantly similar
particles (i.e. finite correlation between particle types). The case
$q=1$ (or $\Delta=\infty$) corresponds to fully identical particle
types, as in the left panel of Fig.~\ref{165328_12Aug17}.

Substituting the above ansatz into the free energy,
Eq.~(\ref{145416_2Aug17}), we obtain
\begin{align}
 \s_m &= \log N + \s_{id}^x + \s_{id}^\sigma + \s_{int},\new
 \s_{id}^x &= -\log\rho -\frac{d}{2}(1-m)\log(2\pi A) + \frac{d}{2}\log m -\frac{d}{2}(1-m),\new
 \s_{id}^{\sigma} & = \log C_m - \frac{\Delta^2}{2}\left(m + m(m-1)q_{m}(\Delta)\right),\new
 \s_{int} &= \frac{\rho}{2}\int d\bm{r} 
 \left[\frac{1}{C_m^2}\int Du Dv Q(\bm{r},u,v)^m -1\right],\new\label{123331_3Aug17}
 \end{align}
where
\begin{align}
  C_m &= \int Du \left[2\cosh(u)\right]^m
\end{align}
and 
\begin{align}
Q(\bm{r},u,v) &= \sum_{\mu\nu}e^{u\sigma(\mu) + v\sigma(\nu)}
 \int d\bm{r}'\gamma_{2A}(\bm{r}+\bm{r}')
 e^{-\beta v_{\mu\nu}(\bm{r}')}.
\end{align}
The order parameters, $A$ and $\Delta$, are determined by the saddle
point conditions, $\partial_A \s_m=0$ and $\partial_\Delta \s_m =0$.
In particular, we focus on the limit of $m\to 1$, which corresponds
to the equilibrium glass transition~\cite{parisi2010}.  
In this limit, we obtain the following self-consistent equations:
\begin{align}
 A &= \frac{1}{\rho}M_A(A,\Delta),\new
 \Delta &= M_\Delta(A,\Delta),\new
 M_A &= \left[-\frac{e^{-\Delta^2}}{4d}\int d\bm{r}\int Du Dv 
 \pdiff{Q}{A}\left[\log Q-f\right]\right]^{-1},\new
 M_\Delta &= -\frac{1}{1+q}\Bigg{[}
-\frac{f'}{2} + \frac{\Delta^2}{2}q' \new
 &+ \frac{\rho \Delta e^{-\Delta^2}}{4}
 \int d\bm{r}DuDv Q\left(\log Q-f\right)\new
 &- \frac{\rho e^{-\Delta^2}}{8}\int d\bm{r} Du' Dv Q\left(\log Q-f\right)\new
 &- \frac{\rho e^{-\Delta^2}}{8}\int d\bm{r} Du Dv' Q\left(\log Q-f\right)\new
 &+ \frac{\rho e^{-\Delta^2}}{8}\int d\bm{r}DuDv Q f'
 \Bigg{]},\label{102818_4Aug17}
\end{align}
where we used the shorthand notation $F'= \partial_\Delta F$
and introduced an auxiliary function;
\begin{align}
 f(\Delta) &= e^{-\Delta^2/2}\int Du \, 2\cosh(u)\log \left[2\cosh(u)\right].
\end{align}

 \subsection{Ansatz for the standard Monte Carlo algorithm}
In case of the standard MC without particle swap,
we assume that  particles of different species can not be
exchanged. All $m$ replicas should have the same species, namely,
\begin{align}
 g(\om) &= \sum_{\mu}x_\mu \prod_{a=1}^m\delta_{\mu^a,\mu},
\end{align}
where $x_\mu=N_\mu/N$ is the number fraction of the $\mu$-species. This
corresponds to the previous ansatz for $\Delta=\infty$.  For
$\rho(\ox)$, we use the same Gaussian ansatz of the swap MC,
Eq.~(\ref{181209_3Aug17}). Substituting the ansatz into the free energy,
Eq.~(\ref{145416_2Aug17}), we obtain
\begin{align}
 \s_m &= \log N + \s_{id} + \s_{int},\new
 \s_{id} &= \log2-\log\rho -\frac{d}{2}(1-m)\log(2\pi A) \\ &+ \frac{d}{2}\log m -\frac{d}{2}(1-m),\new
 \s_{int} &= \frac{\rho}{2}\sum_{\mu\nu}x_\mu x_\nu\int d\bm{r}\left[Q_{\mu\nu}(\bm{r})^m-1\right],
\end{align}
where
\begin{align}
 Q_{\mu\nu}(\bm{r}) &= \int d\bm{r}' \gamma_{2A}(\bm{r}+\bm{r}')e^{-\beta v_{\mu\nu}(\bm{r}')}.
\end{align}
From the saddle point condition $\partial_A \s_m=0$, we can calculate
the value of $A$. In the $m\to 1$ limit, we obtain
\begin{align}
 A &= \frac{1}{\rho}M(A),\label{153245_1Sep17} 
\end{align}
where 
\begin{align}
M(A) &= \left[\frac{1}{d}\sum_{\mu\nu}x_\mu x_\nu\int d\bm{r}Q_{\mu\nu}\log Q_{\mu\nu}\right]^{-1}.\label{102826_4Aug17}
\end{align}

\subsection{Numerical solution of the equations}

The self-consistent equations, Eqs.~(\ref{102818_4Aug17}) and
Eq.~(\ref{153245_1Sep17}), can be solved iteratively. The dynamical
glass transition point $\rho_d$ is defined as the density at which
nontrivial solutions of the order parameters
appear~\cite{parisi2010}. Near $\rho_d$ however, the iterative method
becomes inefficient and it takes a long time to find the transition
point in this way. An efficient way is to calculate the dynamical
transition point from
\begin{align}
 \rho_d^{\rm swap} & = \min_{A}\left[\frac{M_A(A,\Delta(A))}{A}\right],\new
 \rho_d^{\rm  mc} & = \min_{A}\left[\frac{M(A)}{A}\right],\label{181901_4Aug17}
\end{align}
where $\Delta(A)$ is obtained by solving iteratively $\Delta=M_\Delta(A,\Delta)$.

\section{Order parameters and phase diagram}
\label{145947_4Aug17}

In this section, we discuss the density dependence of the order
parameters and the phase diagram obtained by solving the self-consistent
equations derived in Sec.~\ref{180053_3Aug17}.

\begin{figure}[t]
 \includegraphics[width=8cm]{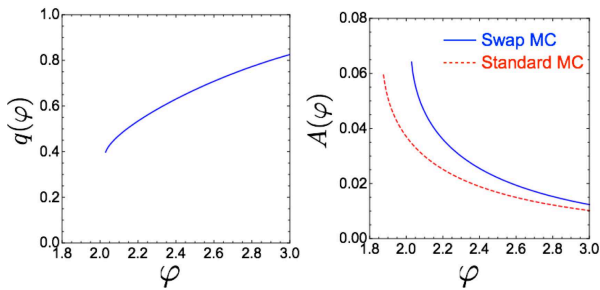}
\caption{\small $\varphi$ dependence of the order parameters at
$\sigma_{LL}/\sigma_{SS}=1.4$ and $d=3$. The blue solid lines denote the
result of the swap MC (with order parameters $q$ and $A$), while the red dashed line denotes the result of
the standard MC (with a single order parameter $A$).}  \label{151657_4Aug17}
\end{figure}

We first discuss the behavior of the swap MC.  We solve
Eqs.~(\ref{102818_4Aug17}) and calculate the order parameters of the
swap MC, $q(\varphi)$ and $A(\varphi)$.  The result at
$\sigma_{LL}/\sigma_{SS}=1.4$ and $d=3$ is shown in
Fig.~\ref{151657_4Aug17} with the blue solid lines.  For sufficiently
small $\varphi$, $A=\infty$ and $\Delta=0$ indicating that there is no
correlation between replicas and the system is in the liquid phase. As
$\varphi$ is increased, $q(\varphi)$ jumps from zero to a finite value
at the dynamical transition point, $\varphi_d^{\rm swap}\approx
2.02$. Simultaneously, $A(\varphi)$ drops from infinity to a finite
value.  This means that, even within the swap MC ansatz, finite
correlations of the particle species spontaneously appear in each glassy
metastable state.  This is consistent with recent computer simulations
where the slowing down of the positional degree of freedom has been
found to be concomitant with that of the
species~\cite{PhysRevX.7.021039}.

For the standard MC, we calculate $A(\varphi)$
by solving Eq.~(\ref{102826_4Aug17}). The result is
shown in the right panel of Fig.~\ref{151657_4Aug17} as a red dashed
line. $A(\varphi)$ of the standard MC changes discontinuously at the
dynamical transition point $\varphi_d^{\rm mc}\approx 1.87$.  The
dynamical transition point of the standard MC is then smaller than that of
the swap MC, $\varphi_d^{\rm mc}<\varphi_d^{\rm swap}$.  This means that
the slow dynamics of the standard MC sets in before that of the swap MC,
providing an explanation for the efficiency of the swap MC.  The value of
$A(\varphi)$ with the swap MC is higher than that of the standard MC,
which is also consistent with recent computer
simulation results~\cite{PhysRevX.7.021039}.

\begin{figure}[t]
 \includegraphics[width=8cm]{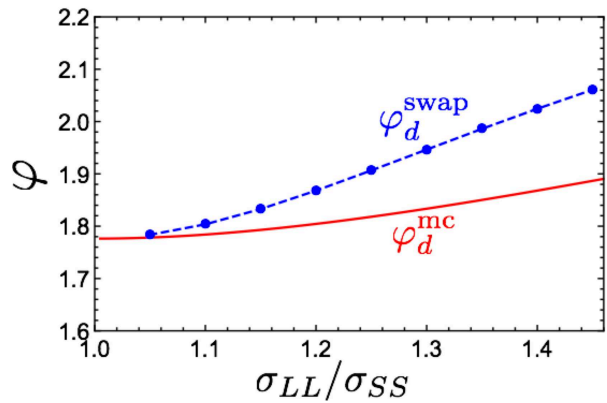}
\caption{\small Phase diagram of the binary MK model. The solid red line
denotes the dynamical transition point of the standard MC. The filled
blue symbols with the dashed line denote the dynamical transition points
of the swap MC. The blue dashed line is an eye guide.}
\label{113732_11Aug17}
\end{figure}

From Eqs.~(\ref{181901_4Aug17}), we can calculate the dynamical
transition point. The results are summarized in
Fig.~\ref{113732_11Aug17}.  As expected, the dynamical transition point of the
standard and swap MC are the same when the size ratio is unity.  The
difference between $\varphi_d^{\rm mc}$ and $\varphi_d^{\rm swap}$
becomes larger with increasing the size ratio.  Note however that, in
Eq.~(\ref{193843_11Aug17}), we assume that the cage size is independent
from the particle species. This assumption would be inappropriate when
the size ratio becomes very large: we discuss in the Appendix a more
general ansatz with two different cages, one for each type of particles.

\section{Computer simulations}
\label{183549_18Aug17}

\subsection{Methods}

In this section, we perform computer simulations of the binary MK model
and compare the results with the theoretical predictions discussed in
Sec.~\ref{145947_4Aug17}.  We employ standard MC and swap MC simulations
for the equimolar binary MK model in $d=3$.  The number of large and
small particles are $N_L=500$ and $N_S=500$, respectively. In case of
the standard MC, we randomly choose a particle and try to shift the
particle position as $(x,y,z)\to (x+\varepsilon r_x, y+\varepsilon
r_y,z+\varepsilon r_z )$, where $r_{\alpha}$ is a random number uniform
in $[-1,1]$ and $\varepsilon$ is an algorithm parameter. We fix
$\varepsilon=0.25$ in this simulation. We accept the new position if the
particle shifted in the new position does not overlap with any other
particle. For the swap MC, in addition to the shift of the particle
position, we try to swap the sizes of particles. We randomly choose two
particles $i$ and $j$, and try to exchange their sizes; note that each
particle keeps its label (i.e. particle $i$ remains $i$ and $j$ remains
$j$) and its random shifts (otherwise the move would never be accepted),
but the sizes of the two particles are exchanged (i.e. particle $i$ now
has the diameter of $j$, and viceversa).  We accept this trial move if
there are no overlapped particles in the new configuration. We try to
shift the positions with probability $1-\alpha$ and to swap the sizes
with probability $\alpha$. As in Ref.~\cite{berthier2016}, we set
$\alpha=0.2$. We prepare initial equilibrated configurations by using
the planting method, which in the MK model allows one to obtain
perfectly equilibrated configurations even beyond
$\varphi_d$~\cite{mari2011,charbonneau2014b}. Below, we report the
results for $\sigma_{LL}/\sigma_{SS}=1.4$. We have confirmed that the
qualitative behavior is unchanged for different values of the size
ratio.

   \subsection{Self-correlation function and relaxation time }
\begin{figure}[t]
 \includegraphics[width=8cm]{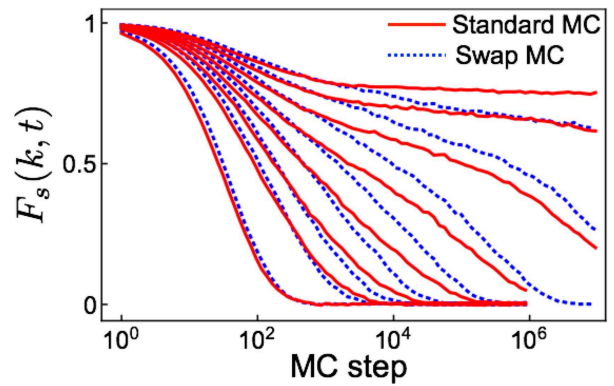}
\caption{\small Equilibrium self-correlation function of the binary MK model.
The red solid lines denote the results of the standard MC, while the
blue dashed lines denote the results of the swap
MC. $\varphi=0.5,1.0,1.2,1.4,1.6,1.8,2.0,2.2$ from left to right.}
\label{170447_11Aug17}
\end{figure}

\begin{figure}[t]
 \includegraphics[width=8cm]{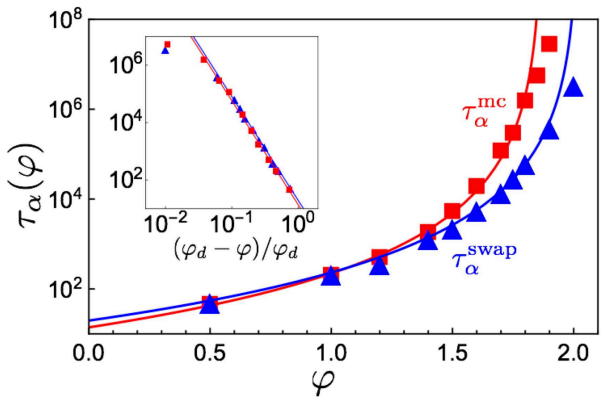}
\caption{\small $\varphi$ dependence of the equilibrium relaxation time
of the binary MK model.  The filled red squares and filled blue
triangles denote the result obtained by the computer simulation of the
standard and swap MC, respectively. The solid lines indicate the
theoretical prediction. (Inset) Scaling plot of the same data of the
main panel.}  \label{190914_11Aug17}
\end{figure}

The first observable that we investigate is the self-correlation
function, which characterizes the slow motion of the particle
positions~\cite{hansen1990}:
\begin{align}
 F_s(k,t) &= \sum_{i=1}^N\frac{1}{N}\left\langle
 e^{i\bm{k}\cdot(\bm{x}_i(t)-\bm{x}_i(0))}\right\rangle,
\end{align}
where $\bm{k}$ denotes the wave vector. Because the system is isotropic,
$F_s(k,t)$ is a function of the absolute value of $\bm{k}$. Following
Ref.~\cite{mari2011}, we set $k=\abs{\bm{k}}= \pi$. In
Fig.~\ref{170447_11Aug17}, we plot $F_s(k,t)$ obtained by the computer
simulation as a function of the MC step, where one MC step is defined as
$N$ MC trials. At very low densities, the relaxation time of the
standard and swap MC are compatible.  On the contrary, at higher
$\varphi$, $F_s(k,t)$ of the swap MC seems to relax faster than that of
the standard MC.
From $F_s(k,t)$, we define the relaxation time, $\tau_\alpha$, by
$F_s(k,\tau_\alpha)=1/e$. The resulting values of $\tau_\alpha$
calculated by the computer simulation are reported in
Fig.~\ref{190914_11Aug17}. To compare with the theoretical prediction,
we fit the numerical data by the power law function predicted by the
MCT~\cite{gotze2009}:
\begin{align}
 \tau_\alpha = \tau_0 (\varphi_d-\varphi)^{-\gamma},\label{192750_11Aug17}
\end{align}
where $\varphi_d$ is not a fitting parameter, as it is 
determined by our theory described in
Sec.~\ref{145947_4Aug17}.  The precise values of $\varphi_d$ for the
standard and swap MC are $\varphi_d^{\rm mc} = 1.87$ and $\varphi_d^{\rm
swap} = 2.02$, respectively.  $\tau_0$ and $\gamma$ are fitting parameters.  The
result of the fitting is shown in Fig.~\ref{190914_11Aug17} with
solid lines. We also show the MCT scaling plot in the inset.  The
scaling formula Eq.~(\ref{192750_11Aug17}) works well for a wide range of
relaxation times, but, for very large values of the relaxation time
($\tau_\alpha>10^6$), the MCT fit systematically overestimates the
relaxation time both for the standard and swap MC.  This is a natural
result because, in finite dimension, the dynamical transition of the MK
model is avoided due to rare hopping events~\cite{charbonneau2014b}.

\subsection{Long time limit of physical quantities}
\begin{figure}[t]
 \includegraphics[width=8cm]{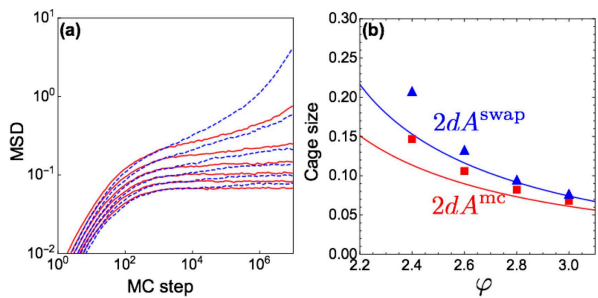}
\caption{\small (a) Equilibrium mean square displacement of the binary
MK model. The red solid and blue dashed lines denote the result of the
standard and swap MC, respectively. $\varphi= 2.0, 2.2, 2.4, 2.6, 2.8,
3.0$ from top to bottom. (b) $\varphi$ dependence of the cage size (long
time limit of the MSD). The filled red squares and blue triangles denote
the numerical results of the standard and swap MC, respectively.  The
theoretical result is shown as a solid line of the same color.}
\label{002754_12Aug17}
\end{figure}

In this subsection, we compare the physical quantities in the long time
limit calculated by theory and computer simulations.  We first
observe the MSD defined by Eq.~\eqref{eq:MSDdef}.
In Fig.~\ref{002754_12Aug17}(a), we show the equilibrium $\msd$ calculated by the
computer simulation.  At small $\varphi$, the $\msd$ continues to grow with
time and diverges in the long time limit. Contrary, at large $\varphi$,
the $\msd$ saturates and converges to a finite value in the long time limit,
meaning that particles are trapped in a narrow region, the ``cage''.  Using the
Gaussian ansatz, Eq.~(\ref{181209_3Aug17}), one can show that the long
time value is related to the cage size, $A$, by
\begin{align}
 \lim_{t\to\infty}\msd(t) &= 2d A.
\end{align}
We approximate the long time limit by the value of $\msd(t)$ at $t=10^7$
and plot the result with the theoretical prediction, see
Fig.~\ref{002754_12Aug17}(b).  The results of the computer simulation
(filled symbols) and theoretical prediction (solid lines) are consistent
at large $\varphi$. However, for small $\varphi$, the theory
underestimates the cage size, as already observed
in~\cite{charbonneau2014b}.  Part of this discrepancy comes from the
poor approximations used in our theory.  In Appendix, we show that one
can obtain a better result by improving the molecular density
approximation, Eq.~(\ref{193843_11Aug17}).

\begin{figure}[t]
 \includegraphics[width=8cm]{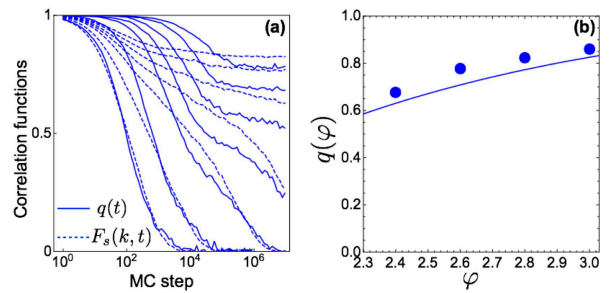}
\caption{\small (a) Equilibrium correlation functions of the swap MC.
The solid line denotes $q(t)$ and the dashed line denotes $F_s(k,t)$.  $\varphi =
 1.0,1.4,1.8,2.0,2.2,2.4,2.6$ from left to right.  (b) $\varphi$
 dependence of $q(\varphi)$. The blue filled circle denotes the numerical
 data. The blue solid line denotes the theoretical prediction.}
\label{152014_12Aug17}
\end{figure}

In the swap MC, the particle species changes with time. The
slow dynamics related to this motion is characterized by
\begin{align}
 q(t) &\equiv \frac{1}{N}\sum_{i=1}^N \left\langle\sigma_i(t)\sigma_i(0)\right\rangle,
\end{align}
where $\sigma_i(t)= 1$ if the $i$-th particle is a large particle at
time $t$, otherwise $\sigma_i(t)=-1$.  In Fig.~\ref{152014_12Aug17}(a),
we show $q(t)$ calculated by computer simulations for several values
of $\varphi$ as a solid line.  We also show $F_s(k,t)$ as a
dashed line. One can see that the relaxation time of $q(t)$ is
comparable to that of $F_s(k,t)$.  This is consistent with the
theoretical prediction that the order parameters of the position and
species begin to have finite values at the same density as shown in
Fig.~\ref{151657_4Aug17}.
Above $\varphi_d^{\rm swap}$, the mean-field theory predicts that $q(t)$
does not decay to zero and converges to a finite value:
\begin{align}
 \lim_{t\to \infty}q(t) = \left\langle\sigma(\mu^a)\sigma(\mu^b) \right\rangle = q.
\end{align}
where $q$ is calculated by the $m\to 1$ limit of
Eq.~(\ref{152834_12Aug17}).  Instead of the long time limit, we evaluate
$q(t)$ at $t=10^7$ MC step and compare with the theoretical prediction.
The result is summarized in Fig.~\ref{152014_12Aug17}(b). The agreement
is good given the simplicity of the approximation. One can
obtain a better result by improving the ansatz, see the Appendix.

\section{Configurational entropy, thermodynamic glass transition point, and activated dynamics}
\label{180943_12Aug17} 

In this section, we discuss the behavior of the configurational entropy,
the thermodynamic glass transition point and the activated dynamics.
Unfortunately, the intensive free energy of the MK model diverges in the
thermodynamic limit, $N\to\infty$, and we can not discuss the
thermodynamic glass transition of this model~\cite{mari2011}. Here
instead, we discuss the qualitative predictions of our theory for more
realistic glass forming systems.  More explicitly, we consider standard
three-dimensional binary or polydisperse mixtures of repulsive
particles, such as the ones discussed in~\cite{PhysRevX.7.021039},
though our discussion may apply to a broader range of systems.

\subsection{Configurational entropy}
\label{sec:conf}
The RFOT scenario and the associated RLT generically (but with notable
exceptions~\cite{berthier2011mic}) predict that the thermodynamic glass
transition point, $\varphi_K$, exists at a higher density than the
dynamic transition density $\varphi_d$.  The configurational entropy
$\con$ characterizes the proximity to the thermodynamic glass transition
point, which is defined by $\con= S_{\rm liq}-S_{\rm glass}$, where
$S_{\rm liq}$ and $S_{\rm glass}$ are the entropies of the liquid and
glass, respectively.  In the liquid phase, $S_{\rm liq}> S_{\rm glass}$
and $\con>0$. With increasing $\varphi$, $\con$ decreases and eventually
vanishes at $\varphi_{K}$.

\begin{figure}[t]
 \includegraphics[width=8cm]{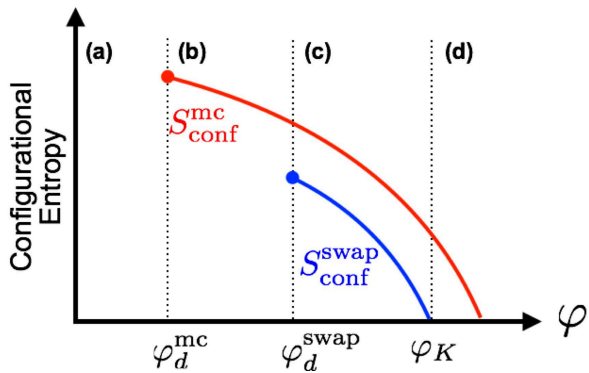}
 \caption{\small Schematic behavior of the configurational entropy.}
 \label{113453_13Aug17}
\end{figure}

However in binary or polydisperse mixtures, our theory provides two
different configurational entropies corresponding to the two different
ansatzes discussed above.  Note that approximate analytical calculations
of $\con$ for realistic three-dimensional systems could be performed
using the two ansatzes, following e.g. the scheme developed
in~\cite{mangeat2016}.  We leave this for future work, and here we limit
ourselves to a schematic discussion of the expected result.  In
Fig.~\ref{113453_13Aug17}, we show the expected behavior of $\con$
calculated by the ansatz corresponding to the standard MC, $\con^{\rm
mc}$, and that corresponding to the swap MC, $\con^{\rm swap}$.
$\con^{\rm mc}$ and $\con^{\rm swap}$ are well defined only above
$\varphi_d^{\rm mc}$ and $\varphi_d^{\rm swap}$, respectively. In
general, $\con^{\rm swap}<\con^{\rm mc}$ because the glass entropy
$S_{\rm glass }$ associated to the swap MC is higher than that of the
standard MC; this is due to the additional degrees of freedom related to
the particle exchange, which is only allowed in the swap MC.  Thus, two
different thermodynamic glass transition points $\varphi_K$ are obtained
from our theory.  The one calculated by $\con^{\rm mc}$ is higher than
that of $\con^{\rm swap}$.

From a purely thermodynamic point of view, the standard MC ansatz does
not have a real meaning. In fact, thermodynamically one seeks to
minimize the free energy over the whole space of functions
$\rho_{\om}(\ox)$, and the swap MC ansatz gives a lower free energy
solution which thus dominates the partition function and the free
energy.  The metastable glassy states in the standard MC ansatz can be
interpreted as an artifact due to the kinetic constraint that prohibits
the exchange of particles of different species.  Thus, the thermodynamic
glass transition point should be determined by $\con^{\rm
swap}(\varphi_K)=0$, or even better, $\varphi_K$ should be determined by
the full optimization of the replicated free energy, i.e. using the most
general ansatz for $\rho_{\om}(\ox)$.

Note that a similar issue in the definition of the configurational
entropy also appears in computer simulation studies.  One should take
into account the exchange of particle species when calculating the
entropy of the glass state (or the vibrational entropy in the
terminology of the computer simulations and experiments), otherwise one
would overestimate the value of the configurational entropy.  The
methods proposed so far seem still inappropriate for this purpose.  For
instance, the inherent structure
method~\cite{scala1999,PhysRevLett.83.3214} and the Frenkel-Ladd
method~\cite{angelani2007} take into account only the vibrational motion
around the equilibrium position and neglect the exchange of the particle
species when calculating the entropy of the glass state. A
generalization of these methods to take into account exchange has been
discussed recently in~\cite{ozawa2017}.  Another approach based on spin
glass theory~\cite{PhysRevLett.79.2486} has been proposed by Berthier
and Coslovich~\cite{berthier2014novel}: using the umbrella sampling,
they calculated the free energy as a function of an overlap order
parameter associated to the particle positions, which partially allows
for particle exchanges. A comparison of the two methods, however, still
reveals discrepancies~\cite{berthier2017breaking}, which might be due to
the approximations involved.  A complete treatment of the
configurational entropy in computer simulations is left for future work.

\subsection{Activated dynamics}
\label{sec:act}

We now discuss the consequences of this structure for the dynamics, but
we warn the reader that the discussion of this subsection is highly
speculative.

From a dynamical point of view, the solution obtained from the standard
MC ansatz might have an important meaning.  The RFOT theory claims that
above the dynamical transition point, the free energy has many
metastable states whose lifetime is controlled by the configurational
entropy~\cite{kirkpatrick1989,bouchaud2004}.  The theory suggests that
in finite dimensional systems, after an initial slowing down controlled
by the MCT scaling in Eq.~\eqref{192750_11Aug17}, the dynamical
transition is avoided and activated dynamics sets in, leading to the
following {\it Adam-Gibbs} relation:
\begin{align}
 \log\tau_{\alpha}\propto \con^{-\eta},\label{111119_28Aug17}
\end{align}
where the critical exponent $\eta$ depends on the shape of the activated
region~\cite{bouchaud2004}.  Eq.~(\ref{111119_28Aug17}) predicts that a
divergence of the relaxation time is concomitant to the vanishing of
$\con$.  For the swap dynamics, the standard RFOT scenario could apply,
using the configurational entropy $\con^{\rm swap}$, with the usual
caveats and limitations discussed extensively
in~\cite{biroli2012random}.

However, in order to apply RFOT arguments to the standard MC, one should
take into account the existence of an additional {\it local} time scale
$\tau_{\rm ex}(T,\varphi)$ that controls particle exchange.  If
$\tau_{\rm ex}$ were to be infinite, then one could apply to the
standard MC the usual RFOT arguments leading to an Adam-Gibbs relation
controlled by $\con^{\rm mc}$.  However, because of its local nature,
the time scale $\tau_{\rm ex}$ cannot diverge at any finite
temperature. Therefore, upon lowering temperature, at some point one
will necessarily have
\begin{equation}
 \log\tau^{\rm mc}_{\alpha}\propto ( \con^{\rm mc})^{-\eta} \gg \log\tau_{\rm ex} \ .
\end{equation}
When this happens, exchange becomes much faster than the lifetime of the metastable states that dominate $\con^{\rm mc}$,
revealing their instability against exchange. 
This argument reveals that there must be a temperature $T_{\rm ex}$ below which (or a density $\varphi_{\rm ex}$ above which)
 $\con^{\rm mc}$ looses its dynamical meaning (we have already seen that it has no thermodynamical meaning).
Below $T_{\rm ex}$ (above $\varphi_{\rm ex}$), the standard MC cannot follow anymore the Adam-Gibbs relation
associated to $\con^{\rm mc}$, and a different dynamics must set in, controlled either by the local exchange processes,
or by the Adam-Gibbs relation associated to the swap dynamics, depending on how the two processes interact. Note that
$T_{\rm ex}$ is expected to be strongly system-dependent due to the local nature of $\tau_{\rm ex}$,  which
depends on the details of the local particle caging. Note also that
$\tau_{\rm ex}$ exists even in the $3d$ MK model~\cite{charbonneau2014b}.

One could well imagine a situation in which $\tau_{\rm ex}$ is small
enough that it destabilizes the whole curve $\con^{\rm mc}$,
i.e. $T_{\rm ex} > T_d^{\rm mc}$ or $\varphi_{\rm ex} < \varphi_d^{\rm
mc}$. In this case, in finite dimensions the finite lifetime of the
states associated to $\con^{\rm mc}$ would be determined by single
particle hopping out of the
cage~\cite{berthier2004length,PhysRevLett.105.135702,PhysRevX.1.021013,charbonneau2014b,ciamarra2016},
rather than by RFOT-like collective phenomena.  In this scenario, one
might therefore expect that the beginning of the slow dynamics in the
region $\varphi_d^{\rm mc} < \varphi < \varphi_d^{\rm swap}$ would be
associated to a local hopping effect, as in the DFT
scenario~\cite{ritort2003,chandler2010,biroli2013}, while around
$\varphi_d^{\rm swap}$ a crossover to the RFOT scenario would be
observed, as also discussed in~\cite{wyart2017does}.  Note that the
precise relation between the hopping kinetic constraint considered above
and the one assumed in the DFT is not clear. Finally, systems with
shorter hopping timescales would exhibit a smaller window of
single-particle slow dynamics before the crossover to the RFOT regime is
reached.  More work is necessary to uncover the precise mechanisms of
the slow dynamics in the region $\varphi_d^{\rm mc} < \varphi <
\varphi_d^{\rm swap}$.

\section{Summary and discussion}
\label{183757_18Aug17}

In this work, we constructed a new ansatz for the replica liquid theory
so as to separately calculate the dynamical glass transition points of
the swap and standard Monte Carlo algorithms within the mean field RFOT
scenario. This is possible by taking into account the effect of the
exchange of particle species. We applied the theory to the binary
Mari-Kurchan (MK) model and calculated the dynamical transition points
of the swap and standard MC, $\varphi_d^{\rm swap}$ and $\varphi_d^{\rm
mc}$, respectively.  We also performed standard and swap MC simulations
of the binary MK model and quantitatively showed that the dynamics in
the standard MC simulation is dominated by $\varphi_d^{\rm mc}$, while
that in the swap MC simulation is dominated by $\varphi_d^{\rm swap}$,
thus validating our ansatzes.

We also discussed qualitatively the thermodynamics and dynamics of more
realistic glass forming systems, as expected from our theory; concrete
calculations could be performed in the future for these systems through
a straightforward extension of the theory.  Four distinct density (or
temperature) regions exist, see Fig.~\ref{113453_13Aug17}.  (a)~When
$\varphi<\varphi_d^{\rm mc}$, the relaxation time of the standard and
swap MC are both small, and the system is liquid.  (b)~When
$\varphi_d^{\rm mc} < \varphi < \varphi_d^{\rm swap}$, there are glassy 
metastable states that are stable only if particle exchange is forbidden. 
Therefore, if one uses the swap MC, the
system relaxes as fast as in the liquid. 
Conversely, if one uses the standard MC, the system displays slow 
dynamics due to the kinetic constraint that
prohibits the exchange of particles species.  
(c)~When $\varphi_d^{\rm swap}<\varphi < \varphi_K$, there are glassy metastable states in the free
energy that remain stable even if particle exchange is allowed.  
At the mean field level, the system is trapped in a metastable
state both for the standard and swap MC, while in finite dimensions the
RFOT scenario should be applicable, and activated relaxation should
dominate the dynamics in both cases.  Note that when
the lifetime of the metastable states 
overcomes the typical time scale to exchange particles of different
species, the relaxation times of
the standard and swap MC would become comparable.  In both cases, the
relaxation time diverges upon approaching $\varphi_K$.  (d)~When
$\varphi>\varphi_K$, the thermodynamic glass transition takes place and
the system loses the ergodicity, remaining arrested in an ideal glass
phase.

It is worth mentioning that, with the appropriate re-scaling, the slow
dynamics of the standard MC is essentially the same as that of other
more realistic dynamics such as the Langevin dynamics, Brownian
dynamics, Newtonian dynamics, and possibly, the true experimental
dynamics~\cite{PhysRevLett.81.4404,szamel2004,berthier2007mon}. Thus,
our results should be translated straightforwardly to these dynamics,
provided the system under investigation is reasonably close to being
mean field, in the sense of a Ginzburg
criterion~\cite{franz2012quantitative}.  The latter property is strongly
system-dependent, and in many systems the mean field scenario can be
heavily affected by finite dimensional fluctuations. In particular, it
is well known that the dynamical transition points $\varphi_d^{\rm mc}$
and $\varphi_d^{\rm swap}$ become simple crossovers in finite
dimensional systems~\cite{biroli2012random}.

Keeping in mind these limitations, our results raise several interesting
points for discussion.
\begin{itemize}
\item[{\it (i)}] In Sec.~\ref{sec:act}, we argued that in one possible
scenario, the exchange time $\tau_{\rm ex}$ is smaller than the
Adam-Gibbs lifetime of the states associated to $\con^{\rm mc}$ already
around $\varphi_d^{\rm mc}$. In this case, $\con^{\rm mc}$ would not
control the slow dynamics. One first observes a slowdown dominated by
the local exchange process, and then a crossover to RFOT-like dynamics
in presence of exchange, which would be associated to $\con^{\rm
swap}$. A similar conclusion has been obtained in the work of Wyart and
Cates~\cite{wyart2017does}. They claim that around the (experimental) glass transition
point $\varphi_g$, the local activation energy $E_{\rm loc}$, which
describes the local physics and cannot diverge, is much larger than the
collective activation energy $E_{\rm cor}$, which is controlled by the
growing static length scale predicted by the RFOT scenario and diverges
at the thermodynamic glass transition point $\varphi_K$ ( or
$T_K$). From this assumption, they concluded that the slow dynamics of
realistic systems is not related to the existence of metastable
states~\cite{wyart2017does}.  We consider that the static length scale
or $E_{\rm cor}$ is controlled by $\con^{\rm swap}$ because $\con^{\rm
mc}$ is meaningless from the thermodynamic point of view as we discussed
in Sec.~\ref{sec:conf}. Thus, Wyart and Cates assumption $E_{\rm loc}\gg
E_{\rm cor}$ is equivalent to assume that $\varphi_g$ is lower than the
density at which a crossover to RFOT-like dynamics associated to
$\con^{\rm swap}$ takes place.

\item[{\it (ii)}] The scenario outlined above, i.e. the fact that the
relaxation of the standard MC dynamics is dominated by local exchange
processes is peculiar, almost by definition, to systems for which the
swap MC dynamics is efficient. In other words, one should keep in mind
that the class of models investigated in~\cite{PhysRevX.7.021039}, for
which the swap algorithm provides a speedup of many orders of magnitude,
could be a specific class of glassy systems for which relaxation is
dominated by local exchange processes. Other glassy systems could behave
differently and present a truly cooperative relaxation. For example, for
one-component or nearly one-component glass forming systems such as the
Gaussian core model~\cite{ikeda2011glass}, the dynamical transition
points of the standard and swap MC are obviously identical and the
region (b) where the slow dynamics is controlled by the kinetic
constraint disappears.  Those models could then display cooperative
relaxation and could thus be ideal playgrounds to test the validity of
the RFOT scenario.
\item[{\it (iii)}] For systems where the swap algorithm is efficient,
and $\varphi_d^{\rm mc} < \varphi_d^{\rm swap}$, within our mean field
framework, we expect that the standard MC dynamics should exhibit a
mode-coupling like phenomenology upon approaching $\varphi_d^{\rm mc}$,
as usual, but the swap MC should also exhibit MCT-like phenomenology
upon approaching $\varphi_d^{\rm swap}$ (of course both transitions
would be avoided in finite dimensions due to activated processes).  In
other words, one expects that the swap MC should develop dynamical
heterogeneities, a critical MCT scaling of the approach to and departure
from the plateau, etc. Some of these phenomena are indeed observed
in~\cite{PhysRevX.7.021039}, but a more systematic study should be
performed.
\item[{\it (iv)}] Because the swap MC should become arrested around
$\varphi_d^{\rm swap}$, i.e. before the metastable states associated to
$\con^{\rm swap}$ are able to develop, this would not be an efficient
algorithm to sample such states.  In particular, the configurational
entropy measured in~\cite{berthier2017breaking} likely pertains to the
region $\varphi_d^{\rm mc} < \varphi < \varphi_d^{\rm swap}$,
i.e. region (b) above, which is the only one accessible to the swap MC.
The configurational entropy $\con^{\rm swap}$ is not well defined in
that region, and therefore its measurement could be plagued by
ambiguities, due to the fact that these states have a finite (and
possibly not so long) lifetime in that region. This is likely to impact
in particular the measurements made via the Frenkel-Ladd method, which
requires states to be stable for long times, while measurement made
through the Franz-Parisi potential should be more
reliable~\cite{berthier2014novel}.
\item[{\it (v)}] Finally, our work shows that even in a region where the
slow dynamics is completely dominated by local kinetic constraints, one
can construct an appropriate thermodynamic theory (in our case, by
forbidding particle exchanges in the construction of replicated
molecules) that is able to capture the associated metastability; a
similar example can be found in Ref.~\cite{foini2012}.
\end{itemize}
We are therefore convinced that our work raises a number of interesting
questions that will hopefully be addressed by future analytical and
numerical works.

\acknowledgments

We thank L.~Berthier, G.~Biroli, J.-P.~Bouchaud, Y.~Jin, K.~Hukushima, T.~Kawasaki, J.~Kurchan, K.~Miyazaki, 
A.~Ninarello, M.~Ozawa, G.~Szamel and H.~Yoshino, for kind
discussions.  This work was supported by a grant from the Simons
Foundation (\#454955, Francesco Zamponi).  A. I. was supported by JSPS
KAKENHI No. 16H04034 and No. 17H04853.  H. I. was supported by JSPS
KAKENHI No. 16J00389.

\widetext
\newpage
\appendix

\section*{Appendix: Two-cage Ansatz}

\begin{figure}[t]
 \includegraphics[width=10cm]{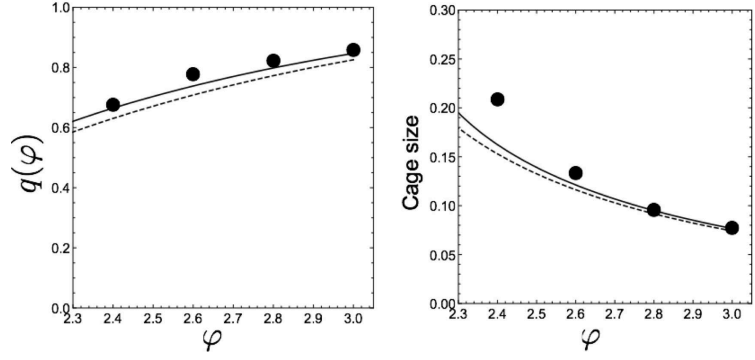}
\caption{\small Order parameters of the swap MC for $\sigma_{LL}/\sigma_{SS}=1.4$. 
The filled symbols denote the results obtained by the computer simulation.
 The solid line denotes the result of the two-cage ansatz. The dashed line denotes the
 result of the one-cage ansatz.}
\label{114912_21Aug17}
\end{figure}

In this Appendix, 
we construct a more general ansatz than
the decoupling approximation in Eq.~(\ref{193843_11Aug17}).
We allow the cage size to depend on the
particles species and make the following ansatz:
\begin{align}
 \rho_{\om}(\ox) &= \rho(\ox|\om)g(\om),
\end{align}
where $g(\om)$ is defined by Eq.~(\ref{201047_2Aug17}) and
\begin{align}
 \rho(\ox|\om) = \rho \int d\bm{X}\prod_{a=1}^m \gamma_{A_{\sigma^a}}(\bm{x}^a-\bm{X})
 = \frac{\rho}{\prod_{a=1}^m \left(2\pi A_{\sigma^a}\right)^{d/2}}
 \left(\frac{2\pi}{\sum_{a=1}^mA_{\sigma^a}^{-1}}\right)^{d/2}
  \exp \left[
 -\frac{1}{4\sum_{a=1}^mA_{\sigma^a}^{-1}}\sum_{ab}
  \frac{(\bm{x}^a-\bm{x}^b)^2}{A_{\sigma^a}  A_{\sigma^b}}
  \right].
\end{align}
Here we used the shorthand notation, $\sigma^a=\sigma(\mu^a)$.  $A_L$
and $A_S$ are the cage sizes of large and small particles, respectively.
Hereafter, we call this the ``two-cage ansatz'', while we refer to
the ansatz in the main text as the ``one-cage ansatz''.  Substituting the
above ansatz into the free energy Eq.~(\ref{145416_2Aug17}), we obtain
\begin{align}
 \frac{\log Z_m}{N} &= S_{id}^{x} + S_{id}^\sigma + S_{int} + \log N,\new
 S_{id}^\sigma &= \log \int Dh \left(2\cosh(h)\right)^m
 -\frac{\Delta^2}{2}\left(m + m(m-1)\frac{\int Dh \cosh(h)^m \tanh(h)^2}{\int Dh \cosh(h)^m}\right),\new
 S_{id}^x &= -\log\rho + \frac{d}{4}m (\log A_L + \log A_S) + (m-1)\frac{d}{2}\log(2\pi)
 + (m-1)\frac{d}{2}\new
 &+ \frac{d}{2}\int_0^\infty \frac{dt}{t}\left[e^{-t}-\frac{1}{\int Dh (2\cosh(h))^m}\int Dh
 \left(e^{h-t/A_L}+e^{-h-t/A_S}\right)^m\right],\new
 S_{int} &= \frac{\rho}{2}\int d\bm{r}\left(\frac{1}{C_m^2}\int Du Dv q(r,u,v)^m-1\right),\new
 q(r,u,v) &= \sum_{\mu\nu}e^{u\sigma(\mu)+v\sigma(\nu)}\int d\bm{u}\gamma_{A_{\sigma(\mu)}+A_{\sigma(\nu)}}(r+u)e^{-\beta v_{\mu\nu}(u)}.
\end{align}
The order parameters are calculated by the saddle point conditions,
$\partial_{A_L}\log Z_m = 0$, $\partial_{A_S}\log Z_m=0$, and
$\partial_{\Delta}\log Z_m=0$. After some manipulations, we obtain
the following self-consistent equations:
\begin{align}
 A_L &= M_L(A_L,A_S,\Delta),\new
 A_S &= M_S(A_L,A_S,\Delta),\new
 \Delta &= M_\Delta(A_L,A_S,\Delta),
\end{align}
where
\begin{align}
 M_L(A_L,A_S,\Delta) &= \frac{\frac{d}{4}
 + \frac{d}{2}\int_0^\infty \frac{dt}{A_L} e^{-t/A_L}\left[\frac{f}{4}-\frac{K_L}{2}\right]}
 {-\frac{\rho}{8}e^{-\Delta^2}\int dr DhDh' \pdiff{q}{A_L}\left(\log q -f \right)},\new
  M_S(A_L,A_S,\Delta) &= \frac{\frac{d}{4}
 + \frac{d}{2}\int_0^\infty\frac{dt}{A_S} e^{-t/A_S}\left[\frac{f}{4}-\frac{K_S}{2}\right]}
 {-\frac{\rho}{8}e^{-\Delta^2}\int dr DhDh' \pdiff{q}{A_S}\left(\log q -f \right)},\new
 M_\Delta(A_L,A_S,\Delta) &= -\frac{1}{1+q}\Bigg{[}
-\frac{f'}{2} + \frac{\Delta^2}{2}g'
 -\frac{d}{2}\int \frac{dt}{t}\left[\frac{f'}{2}\frac{e^{-t/A_l}+e^{-t/A_S}}{2}\frac{K'}{2}\right]\new
 &+ \frac{\rho}{4}\Delta e^{-\Delta^2}\int q(\log q-f)-\frac{\rho}{8}e^{-\Delta^2}
 \int dr \pdiff{Dh}{\Delta}Dh' q(\log q-f)
 -\frac{\rho}{8}e^{-\Delta^2}
 \int dr Dh\pdiff{Dh'}{\Delta} q(\log q-f)\new
 &+ \frac{\rho}{8}e^{-\Delta^2}\int q f' \Bigg{]}.
\end{align}
We have introduced the auxiliary functions, $K$, $K_L$ and $K_S$ as
\begin{align}
 K(A_L,A_S,\Delta) &= e^{-\Delta^2/2}\int Dh \left(e^{h-t/A_L}+e^{-h-t/A_S}\right)\log \left(e^{h-t/A_L}+e^{-h-t/A_S}\right),\new
K_L(A_L,A_A,\Delta) &= 1 + e^{-\Delta^2/2}\int Dh e^h \log \left(e^{h-t/A_L}+e^{-h-t/A_S}\right),\new
K_S(A_L,A_S,\Delta) &= 
 1 + e^{-\Delta^2/2}\int Dh e^{-h} \log \left(e^{h-t/A_L}+e^{-h-t/A_S}\right).
\end{align}
We solved the self-consistent equations by using the iterative method
for the size ratio $\sigma_{LL}/\sigma_{SS}=1.4$.  The result is
summarized in Fig.~\ref{114912_21Aug17}. One can see that the two cage
ansatz gives a slightly better result than the one-cage ansatz at this
size ratio.  Our preliminary calculations for smaller size ratio
$\sigma_{LL}/\sigma_{SS}\approx 1.2$ predict, however, a decoupling of
the glass transition point of the position and species, which was never
observed in computer simulation.  The reason for this discrepancy
between the theory and computer simulations is still unclear and its
clarification is left for future work.

\bibliography{reference}

\begin{thebibliography}{74}%
\makeatletter
\providecommand \@ifxundefined [1]{%
 \@ifx{#1\undefined}
}%
\providecommand \@ifnum [1]{%
 \ifnum #1\expandafter \@firstoftwo
 \else \expandafter \@secondoftwo
 \fi
}%
\providecommand \@ifx [1]{%
 \ifx #1\expandafter \@firstoftwo
 \else \expandafter \@secondoftwo
 \fi
}%
\providecommand \natexlab [1]{#1}%
\providecommand \enquote  [1]{``#1''}%
\providecommand \bibnamefont  [1]{#1}%
\providecommand \bibfnamefont [1]{#1}%
\providecommand \citenamefont [1]{#1}%
\providecommand \href@noop [0]{\@secondoftwo}%
\providecommand \href [0]{\begingroup \@sanitize@url \@href}%
\providecommand \@href[1]{\@@startlink{#1}\@@href}%
\providecommand \@@href[1]{\endgroup#1\@@endlink}%
\providecommand \@sanitize@url [0]{\catcode `\\12\catcode `\$12\catcode
  `\&12\catcode `\#12\catcode `\^12\catcode `\_12\catcode `\%12\relax}%
\providecommand \@@startlink[1]{}%
\providecommand \@@endlink[0]{}%
\providecommand \url  [0]{\begingroup\@sanitize@url \@url }%
\providecommand \@url [1]{\endgroup\@href {#1}{\urlprefix }}%
\providecommand \urlprefix  [0]{URL }%
\providecommand \Eprint [0]{\href }%
\providecommand \doibase [0]{http://dx.doi.org/}%
\providecommand \selectlanguage [0]{\@gobble}%
\providecommand \bibinfo  [0]{\@secondoftwo}%
\providecommand \bibfield  [0]{\@secondoftwo}%
\providecommand \translation [1]{[#1]}%
\providecommand \BibitemOpen [0]{}%
\providecommand \bibitemStop [0]{}%
\providecommand \bibitemNoStop [0]{.\EOS\space}%
\providecommand \EOS [0]{\spacefactor3000\relax}%
\providecommand \BibitemShut  [1]{\csname bibitem#1\endcsname}%
\let\auto@bib@innerbib\@empty
\bibitem [{\citenamefont {Debenedetti}\ and\ \citenamefont
  {Stillinger}(2001)}]{debene2001}%
  \BibitemOpen
  \bibfield  {author} {\bibinfo {author} {\bibfnamefont {P.~G.}\ \bibnamefont
  {Debenedetti}}\ and\ \bibinfo {author} {\bibfnamefont {F.~H.}\ \bibnamefont
  {Stillinger}},\ }\href@noop {} {\bibfield  {journal} {\bibinfo  {journal}
  {Nat.}\ }\textbf {\bibinfo {volume} {410}},\ \bibinfo {pages} {259} (\bibinfo
  {year} {2001})}\BibitemShut {NoStop}%
\bibitem [{\citenamefont {Cavagna}(2009)}]{cavagna2009}%
  \BibitemOpen
  \bibfield  {author} {\bibinfo {author} {\bibfnamefont {A.}~\bibnamefont
  {Cavagna}},\ }\href@noop {} {\bibfield  {journal} {\bibinfo  {journal} {Phys.
  Rep.}\ }\textbf {\bibinfo {volume} {476}},\ \bibinfo {pages} {51} (\bibinfo
  {year} {2009})}\BibitemShut {NoStop}%
\bibitem [{\citenamefont {Gotze}(2009)}]{gotze2009}%
  \BibitemOpen
  \bibfield  {author} {\bibinfo {author} {\bibfnamefont {W.}~\bibnamefont
  {Gotze}},\ }\href@noop {} {\emph {\bibinfo {title} {Complex dynamics of
  glass-forming liquids}}}\ (\bibinfo  {publisher} {Oxford University Press},\
  \bibinfo {year} {2009})\BibitemShut {NoStop}%
\bibitem [{\citenamefont {Berthier}\ and\ \citenamefont
  {Biroli}(2011)}]{berthier2011}%
  \BibitemOpen
  \bibfield  {author} {\bibinfo {author} {\bibfnamefont {L.}~\bibnamefont
  {Berthier}}\ and\ \bibinfo {author} {\bibfnamefont {G.}~\bibnamefont
  {Biroli}},\ }\href@noop {} {\bibfield  {journal} {\bibinfo  {journal} {Rev.
  Mod. Phys.}\ }\textbf {\bibinfo {volume} {83}},\ \bibinfo {pages} {587}
  (\bibinfo {year} {2011})}\BibitemShut {NoStop}%
\bibitem [{\citenamefont {Biroli}\ and\ \citenamefont
  {Garrahan}(2013)}]{biroli2013}%
  \BibitemOpen
  \bibfield  {author} {\bibinfo {author} {\bibfnamefont {G.}~\bibnamefont
  {Biroli}}\ and\ \bibinfo {author} {\bibfnamefont {J.~P.}\ \bibnamefont
  {Garrahan}},\ }\href@noop {} {\bibfield  {journal} {\bibinfo  {journal} {J.
  Chem. Phys.}\ }\textbf {\bibinfo {volume} {138}},\ \bibinfo {pages} {12A301}
  (\bibinfo {year} {2013})}\BibitemShut {NoStop}%
\bibitem [{\citenamefont {Kirkpatrick}\ and\ \citenamefont
  {Wolynes}(1987)}]{kirkpatrick1987}%
  \BibitemOpen
  \bibfield  {author} {\bibinfo {author} {\bibfnamefont {T.}~\bibnamefont
  {Kirkpatrick}}\ and\ \bibinfo {author} {\bibfnamefont {P.}~\bibnamefont
  {Wolynes}},\ }\href@noop {} {\bibfield  {journal} {\bibinfo  {journal} {Phys.
  Rev. A}\ }\textbf {\bibinfo {volume} {35}},\ \bibinfo {pages} {3072}
  (\bibinfo {year} {1987})}\BibitemShut {NoStop}%
\bibitem [{\citenamefont {Kirkpatrick}\ \emph {et~al.}(1989)\citenamefont
  {Kirkpatrick}, \citenamefont {Thirumalai},\ and\ \citenamefont
  {Wolynes}}]{kirkpatrick1989}%
  \BibitemOpen
  \bibfield  {author} {\bibinfo {author} {\bibfnamefont {T.}~\bibnamefont
  {Kirkpatrick}}, \bibinfo {author} {\bibfnamefont {D.}~\bibnamefont
  {Thirumalai}}, \ and\ \bibinfo {author} {\bibfnamefont {P.~G.}\ \bibnamefont
  {Wolynes}},\ }\href@noop {} {\bibfield  {journal} {\bibinfo  {journal} {Phys.
  Rev. A}\ }\textbf {\bibinfo {volume} {40}},\ \bibinfo {pages} {1045}
  (\bibinfo {year} {1989})}\BibitemShut {NoStop}%
\bibitem [{\citenamefont {Bouchaud}\ and\ \citenamefont
  {Biroli}(2004)}]{bouchaud2004}%
  \BibitemOpen
  \bibfield  {author} {\bibinfo {author} {\bibfnamefont {J.-P.}\ \bibnamefont
  {Bouchaud}}\ and\ \bibinfo {author} {\bibfnamefont {G.}~\bibnamefont
  {Biroli}},\ }\href@noop {} {\bibfield  {journal} {\bibinfo  {journal} {J.
  Chem. Phys.}\ }\textbf {\bibinfo {volume} {121}},\ \bibinfo {pages} {7347}
  (\bibinfo {year} {2004})}\BibitemShut {NoStop}%
\bibitem [{\citenamefont {Lubchenko}\ and\ \citenamefont
  {Wolynes}(2007)}]{lubchenko2007}%
  \BibitemOpen
  \bibfield  {author} {\bibinfo {author} {\bibfnamefont {V.}~\bibnamefont
  {Lubchenko}}\ and\ \bibinfo {author} {\bibfnamefont {P.~G.}\ \bibnamefont
  {Wolynes}},\ }\href@noop {} {\bibfield  {journal} {\bibinfo  {journal} {Annu.
  Rev. Phys. Chem.}\ }\textbf {\bibinfo {volume} {58}},\ \bibinfo {pages} {235}
  (\bibinfo {year} {2007})}\BibitemShut {NoStop}%
\bibitem [{\citenamefont {Kirkpatrick}\ and\ \citenamefont
  {Thirumalai}(2015)}]{kirkpatrick2015}%
  \BibitemOpen
  \bibfield  {author} {\bibinfo {author} {\bibfnamefont {T.~R.}\ \bibnamefont
  {Kirkpatrick}}\ and\ \bibinfo {author} {\bibfnamefont {D.}~\bibnamefont
  {Thirumalai}},\ }\href@noop {} {\bibfield  {journal} {\bibinfo  {journal}
  {Rev. Mod. Phys.}\ }\textbf {\bibinfo {volume} {87}},\ \bibinfo {pages} {183}
  (\bibinfo {year} {2015})}\BibitemShut {NoStop}%
\bibitem [{\citenamefont {Parisi}\ and\ \citenamefont
  {Zamponi}(2010)}]{parisi2010}%
  \BibitemOpen
  \bibfield  {author} {\bibinfo {author} {\bibfnamefont {G.}~\bibnamefont
  {Parisi}}\ and\ \bibinfo {author} {\bibfnamefont {F.}~\bibnamefont
  {Zamponi}},\ }\href@noop {} {\bibfield  {journal} {\bibinfo  {journal} {Rev.
  Mod. Phys.}\ }\textbf {\bibinfo {volume} {82}},\ \bibinfo {pages} {789}
  (\bibinfo {year} {2010})}\BibitemShut {NoStop}%
\bibitem [{\citenamefont {Charbonneau}\ \emph {et~al.}(2017)\citenamefont
  {Charbonneau}, \citenamefont {Kurchan}, \citenamefont {Parisi}, \citenamefont
  {Urbani},\ and\ \citenamefont {Zamponi}}]{charbonneau2017}%
  \BibitemOpen
  \bibfield  {author} {\bibinfo {author} {\bibfnamefont {P.}~\bibnamefont
  {Charbonneau}}, \bibinfo {author} {\bibfnamefont {J.}~\bibnamefont
  {Kurchan}}, \bibinfo {author} {\bibfnamefont {G.}~\bibnamefont {Parisi}},
  \bibinfo {author} {\bibfnamefont {P.}~\bibnamefont {Urbani}}, \ and\ \bibinfo
  {author} {\bibfnamefont {F.}~\bibnamefont {Zamponi}},\ }\href@noop {}
  {\bibfield  {journal} {\bibinfo  {journal} {Annu. Rev. Condens. Matter
  Phys.}\ }\textbf {\bibinfo {volume} {8}},\ \bibinfo {pages} {265} (\bibinfo
  {year} {2017})}\BibitemShut {NoStop}%
\bibitem [{\citenamefont {M{\'e}zard}\ and\ \citenamefont
  {Parisi}(1999{\natexlab{a}})}]{mezard1999}%
  \BibitemOpen
  \bibfield  {author} {\bibinfo {author} {\bibfnamefont {M.}~\bibnamefont
  {M{\'e}zard}}\ and\ \bibinfo {author} {\bibfnamefont {G.}~\bibnamefont
  {Parisi}},\ }\href@noop {} {\bibfield  {journal} {\bibinfo  {journal} {J.
  Phys. Condens. Matter}\ }\textbf {\bibinfo {volume} {11}},\ \bibinfo {pages}
  {A157} (\bibinfo {year} {1999}{\natexlab{a}})}\BibitemShut {NoStop}%
\bibitem [{\citenamefont {M{\'e}zard}\ and\ \citenamefont
  {Parisi}(1999{\natexlab{b}})}]{mezard1999f}%
  \BibitemOpen
  \bibfield  {author} {\bibinfo {author} {\bibfnamefont {M.}~\bibnamefont
  {M{\'e}zard}}\ and\ \bibinfo {author} {\bibfnamefont {G.}~\bibnamefont
  {Parisi}},\ }\href@noop {} {\bibfield  {journal} {\bibinfo  {journal} {J.
  Chem. Phys.}\ }\textbf {\bibinfo {volume} {111}},\ \bibinfo {pages} {1076}
  (\bibinfo {year} {1999}{\natexlab{b}})}\BibitemShut {NoStop}%
\bibitem [{\citenamefont {Parisi}\ and\ \citenamefont
  {Zamponi}(2005)}]{parisi2005}%
  \BibitemOpen
  \bibfield  {author} {\bibinfo {author} {\bibfnamefont {G.}~\bibnamefont
  {Parisi}}\ and\ \bibinfo {author} {\bibfnamefont {F.}~\bibnamefont
  {Zamponi}},\ }\href@noop {} {\bibfield  {journal} {\bibinfo  {journal} {J.
  Chem. Phys.}\ }\textbf {\bibinfo {volume} {123}},\ \bibinfo {pages} {144501}
  (\bibinfo {year} {2005})}\BibitemShut {NoStop}%
\bibitem [{\citenamefont {Jacquin}\ \emph {et~al.}(2011)\citenamefont
  {Jacquin}, \citenamefont {Berthier},\ and\ \citenamefont
  {Zamponi}}]{PhysRevLett.106.135702}%
  \BibitemOpen
  \bibfield  {author} {\bibinfo {author} {\bibfnamefont {H.}~\bibnamefont
  {Jacquin}}, \bibinfo {author} {\bibfnamefont {L.}~\bibnamefont {Berthier}}, \
  and\ \bibinfo {author} {\bibfnamefont {F.}~\bibnamefont {Zamponi}},\ }\href
  {\doibase 10.1103/PhysRevLett.106.135702} {\bibfield  {journal} {\bibinfo
  {journal} {Phys. Rev. Lett.}\ }\textbf {\bibinfo {volume} {106}},\ \bibinfo
  {pages} {135702} (\bibinfo {year} {2011})}\BibitemShut {NoStop}%
\bibitem [{\citenamefont {Berthier}\ \emph {et~al.}(2011)\citenamefont
  {Berthier}, \citenamefont {Jacquin},\ and\ \citenamefont
  {Zamponi}}]{berthier2011mic}%
  \BibitemOpen
  \bibfield  {author} {\bibinfo {author} {\bibfnamefont {L.}~\bibnamefont
  {Berthier}}, \bibinfo {author} {\bibfnamefont {H.}~\bibnamefont {Jacquin}}, \
  and\ \bibinfo {author} {\bibfnamefont {F.}~\bibnamefont {Zamponi}},\
  }\href@noop {} {\bibfield  {journal} {\bibinfo  {journal} {Phys. Rev. E}\
  }\textbf {\bibinfo {volume} {84}},\ \bibinfo {pages} {051103} (\bibinfo
  {year} {2011})}\BibitemShut {NoStop}%
\bibitem [{\citenamefont {Mangeat}\ and\ \citenamefont
  {Zamponi}(2016)}]{mangeat2016}%
  \BibitemOpen
  \bibfield  {author} {\bibinfo {author} {\bibfnamefont {M.}~\bibnamefont
  {Mangeat}}\ and\ \bibinfo {author} {\bibfnamefont {F.}~\bibnamefont
  {Zamponi}},\ }\href@noop {} {\bibfield  {journal} {\bibinfo  {journal} {Phys
  Rev. E}\ }\textbf {\bibinfo {volume} {93}},\ \bibinfo {pages} {012609}
  (\bibinfo {year} {2016})}\BibitemShut {NoStop}%
\bibitem [{\citenamefont {Bengtzelius}\ \emph {et~al.}(1984)\citenamefont
  {Bengtzelius}, \citenamefont {Gotze},\ and\ \citenamefont
  {Sjolander}}]{bengtzelius1984}%
  \BibitemOpen
  \bibfield  {author} {\bibinfo {author} {\bibfnamefont {U.}~\bibnamefont
  {Bengtzelius}}, \bibinfo {author} {\bibfnamefont {W.}~\bibnamefont {Gotze}},
  \ and\ \bibinfo {author} {\bibfnamefont {A.}~\bibnamefont {Sjolander}},\
  }\href@noop {} {\bibfield  {journal} {\bibinfo  {journal} {J. Phys. C}\
  }\textbf {\bibinfo {volume} {17}},\ \bibinfo {pages} {5915} (\bibinfo {year}
  {1984})}\BibitemShut {NoStop}%
\bibitem [{\citenamefont {Maimbourg}\ \emph {et~al.}(2016)\citenamefont
  {Maimbourg}, \citenamefont {Kurchan},\ and\ \citenamefont
  {Zamponi}}]{maim2016}%
  \BibitemOpen
  \bibfield  {author} {\bibinfo {author} {\bibfnamefont {T.}~\bibnamefont
  {Maimbourg}}, \bibinfo {author} {\bibfnamefont {J.}~\bibnamefont {Kurchan}},
  \ and\ \bibinfo {author} {\bibfnamefont {F.}~\bibnamefont {Zamponi}},\ }\href
  {\doibase 10.1103/PhysRevLett.116.015902} {\bibfield  {journal} {\bibinfo
  {journal} {Phys. Rev. Lett.}\ }\textbf {\bibinfo {volume} {116}},\ \bibinfo
  {pages} {015902} (\bibinfo {year} {2016})}\BibitemShut {NoStop}%
\bibitem [{\citenamefont {Adam}\ and\ \citenamefont {Gibbs}(1965)}]{adam1965}%
  \BibitemOpen
  \bibfield  {author} {\bibinfo {author} {\bibfnamefont {G.}~\bibnamefont
  {Adam}}\ and\ \bibinfo {author} {\bibfnamefont {J.~H.}\ \bibnamefont
  {Gibbs}},\ }\href@noop {} {\bibfield  {journal} {\bibinfo  {journal} {J.
  Chem. Phys.}\ }\textbf {\bibinfo {volume} {43}},\ \bibinfo {pages} {139}
  (\bibinfo {year} {1965})}\BibitemShut {NoStop}%
\bibitem [{\citenamefont {Kauzmann}(1948)}]{kauzmann1948}%
  \BibitemOpen
  \bibfield  {author} {\bibinfo {author} {\bibfnamefont {W.}~\bibnamefont
  {Kauzmann}},\ }\href@noop {} {\bibfield  {journal} {\bibinfo  {journal}
  {Chem. Rev.}\ }\textbf {\bibinfo {volume} {43}},\ \bibinfo {pages} {219}
  (\bibinfo {year} {1948})}\BibitemShut {NoStop}%
\bibitem [{\citenamefont {Biroli}\ \emph {et~al.}()\citenamefont {Biroli},
  \citenamefont {Bouchaud}, \citenamefont {Cavagna}, \citenamefont {Grigera},\
  and\ \citenamefont {Verrocchio}}]{biroli2008}%
  \BibitemOpen
  \bibfield  {author} {\bibinfo {author} {\bibfnamefont {G.}~\bibnamefont
  {Biroli}}, \bibinfo {author} {\bibfnamefont {J.-P.}\ \bibnamefont
  {Bouchaud}}, \bibinfo {author} {\bibfnamefont {A.}~\bibnamefont {Cavagna}},
  \bibinfo {author} {\bibfnamefont {T.~S.}\ \bibnamefont {Grigera}}, \ and\
  \bibinfo {author} {\bibfnamefont {P.}~\bibnamefont {Verrocchio}},\
  }\href@noop {} {\bibfield  {journal} {\bibinfo  {journal} {Nat. Phys.}\
  }\textbf {\bibinfo {volume} {4}},\ \bibinfo {pages} {771}}\BibitemShut
  {NoStop}%
\bibitem [{\citenamefont {Berthier}\ and\ \citenamefont
  {Kob}(2012)}]{berthier2012static}%
  \BibitemOpen
  \bibfield  {author} {\bibinfo {author} {\bibfnamefont {L.}~\bibnamefont
  {Berthier}}\ and\ \bibinfo {author} {\bibfnamefont {W.}~\bibnamefont {Kob}},\
  }\href@noop {} {\bibfield  {journal} {\bibinfo  {journal} {Phys. Rev. E}\
  }\textbf {\bibinfo {volume} {85}},\ \bibinfo {pages} {011102} (\bibinfo
  {year} {2012})}\BibitemShut {NoStop}%
\bibitem [{\citenamefont {Cammarota}\ and\ \citenamefont
  {Biroli}(2012)}]{cammarota2012ideal}%
  \BibitemOpen
  \bibfield  {author} {\bibinfo {author} {\bibfnamefont {C.}~\bibnamefont
  {Cammarota}}\ and\ \bibinfo {author} {\bibfnamefont {G.}~\bibnamefont
  {Biroli}},\ }\href@noop {} {\bibfield  {journal} {\bibinfo  {journal} {PNAS}\
  }\textbf {\bibinfo {volume} {109}},\ \bibinfo {pages} {8850} (\bibinfo {year}
  {2012})}\BibitemShut {NoStop}%
\bibitem [{\citenamefont {Cammarota}\ and\ \citenamefont
  {Biroli}(2013)}]{cammarota2013random}%
  \BibitemOpen
  \bibfield  {author} {\bibinfo {author} {\bibfnamefont {C.}~\bibnamefont
  {Cammarota}}\ and\ \bibinfo {author} {\bibfnamefont {G.}~\bibnamefont
  {Biroli}},\ }\href@noop {} {\bibfield  {journal} {\bibinfo  {journal} {J.
  Chem. Phys.}\ }\textbf {\bibinfo {volume} {138}},\ \bibinfo {pages} {12A547}
  (\bibinfo {year} {2013})}\BibitemShut {NoStop}%
\bibitem [{\citenamefont {Karmakar}\ and\ \citenamefont
  {Parisi}(2013)}]{karmakar2013random}%
  \BibitemOpen
  \bibfield  {author} {\bibinfo {author} {\bibfnamefont {S.}~\bibnamefont
  {Karmakar}}\ and\ \bibinfo {author} {\bibfnamefont {G.}~\bibnamefont
  {Parisi}},\ }\href@noop {} {\bibfield  {journal} {\bibinfo  {journal} {PNAS}\
  }\textbf {\bibinfo {volume} {110}},\ \bibinfo {pages} {2752} (\bibinfo {year}
  {2013})}\BibitemShut {NoStop}%
\bibitem [{\citenamefont {Kob}\ and\ \citenamefont
  {Berthier}(2013)}]{kob2013probing}%
  \BibitemOpen
  \bibfield  {author} {\bibinfo {author} {\bibfnamefont {W.}~\bibnamefont
  {Kob}}\ and\ \bibinfo {author} {\bibfnamefont {L.}~\bibnamefont {Berthier}},\
  }\href@noop {} {\bibfield  {journal} {\bibinfo  {journal} {Phys. Rev. Lett.}\
  }\textbf {\bibinfo {volume} {110}},\ \bibinfo {pages} {245702} (\bibinfo
  {year} {2013})}\BibitemShut {NoStop}%
\bibitem [{\citenamefont {Ozawa}\ \emph {et~al.}(2015)\citenamefont {Ozawa},
  \citenamefont {Kob}, \citenamefont {Ikeda},\ and\ \citenamefont
  {Miyazaki}}]{ozawa2015equilibrium}%
  \BibitemOpen
  \bibfield  {author} {\bibinfo {author} {\bibfnamefont {M.}~\bibnamefont
  {Ozawa}}, \bibinfo {author} {\bibfnamefont {W.}~\bibnamefont {Kob}}, \bibinfo
  {author} {\bibfnamefont {A.}~\bibnamefont {Ikeda}}, \ and\ \bibinfo {author}
  {\bibfnamefont {K.}~\bibnamefont {Miyazaki}},\ }\href@noop {} {\bibfield
  {journal} {\bibinfo  {journal} {PNAS}\ }\textbf {\bibinfo {volume} {112}},\
  \bibinfo {pages} {6914} (\bibinfo {year} {2015})}\BibitemShut {NoStop}%
\bibitem [{\citenamefont {Ritort}\ and\ \citenamefont
  {Sollich}(2003)}]{ritort2003}%
  \BibitemOpen
  \bibfield  {author} {\bibinfo {author} {\bibfnamefont {F.}~\bibnamefont
  {Ritort}}\ and\ \bibinfo {author} {\bibfnamefont {P.}~\bibnamefont
  {Sollich}},\ }\href@noop {} {\bibfield  {journal} {\bibinfo  {journal} {Adv.
  Phys.}\ }\textbf {\bibinfo {volume} {52}},\ \bibinfo {pages} {219} (\bibinfo
  {year} {2003})}\BibitemShut {NoStop}%
\bibitem [{\citenamefont {Chandler}\ and\ \citenamefont
  {Garrahan}(2010)}]{chandler2010}%
  \BibitemOpen
  \bibfield  {author} {\bibinfo {author} {\bibfnamefont {D.}~\bibnamefont
  {Chandler}}\ and\ \bibinfo {author} {\bibfnamefont {J.~P.}\ \bibnamefont
  {Garrahan}},\ }\href@noop {} {\bibfield  {journal} {\bibinfo  {journal}
  {Annu. Rev. Phys. Chem.}\ }\textbf {\bibinfo {volume} {61}},\ \bibinfo
  {pages} {191} (\bibinfo {year} {2010})}\BibitemShut {NoStop}%
\bibitem [{\citenamefont {Keys}\ \emph {et~al.}(2011)\citenamefont {Keys},
  \citenamefont {Hedges}, \citenamefont {Garrahan}, \citenamefont {Glotzer},\
  and\ \citenamefont {Chandler}}]{PhysRevX.1.021013}%
  \BibitemOpen
  \bibfield  {author} {\bibinfo {author} {\bibfnamefont {A.~S.}\ \bibnamefont
  {Keys}}, \bibinfo {author} {\bibfnamefont {L.~O.}\ \bibnamefont {Hedges}},
  \bibinfo {author} {\bibfnamefont {J.~P.}\ \bibnamefont {Garrahan}}, \bibinfo
  {author} {\bibfnamefont {S.~C.}\ \bibnamefont {Glotzer}}, \ and\ \bibinfo
  {author} {\bibfnamefont {D.}~\bibnamefont {Chandler}},\ }\href {\doibase
  10.1103/PhysRevX.1.021013} {\bibfield  {journal} {\bibinfo  {journal} {Phys.
  Rev. X}\ }\textbf {\bibinfo {volume} {1}},\ \bibinfo {pages} {021013}
  (\bibinfo {year} {2011})}\BibitemShut {NoStop}%
\bibitem [{\citenamefont {Isobe}\ \emph {et~al.}(2016)\citenamefont {Isobe},
  \citenamefont {Keys}, \citenamefont {Chandler},\ and\ \citenamefont
  {Garrahan}}]{PhysRevLett.117.145701}%
  \BibitemOpen
  \bibfield  {author} {\bibinfo {author} {\bibfnamefont {M.}~\bibnamefont
  {Isobe}}, \bibinfo {author} {\bibfnamefont {A.~S.}\ \bibnamefont {Keys}},
  \bibinfo {author} {\bibfnamefont {D.}~\bibnamefont {Chandler}}, \ and\
  \bibinfo {author} {\bibfnamefont {J.~P.}\ \bibnamefont {Garrahan}},\ }\href
  {\doibase 10.1103/PhysRevLett.117.145701} {\bibfield  {journal} {\bibinfo
  {journal} {Phys. Rev. Lett.}\ }\textbf {\bibinfo {volume} {117}},\ \bibinfo
  {pages} {145701} (\bibinfo {year} {2016})}\BibitemShut {NoStop}%
\bibitem [{\citenamefont {Fredrickson}\ and\ \citenamefont
  {Andersen}(1984)}]{PhysRevLett.53.1244}%
  \BibitemOpen
  \bibfield  {author} {\bibinfo {author} {\bibfnamefont {G.~H.}\ \bibnamefont
  {Fredrickson}}\ and\ \bibinfo {author} {\bibfnamefont {H.~C.}\ \bibnamefont
  {Andersen}},\ }\href {\doibase 10.1103/PhysRevLett.53.1244} {\bibfield
  {journal} {\bibinfo  {journal} {Phys. Rev. Lett.}\ }\textbf {\bibinfo
  {volume} {53}},\ \bibinfo {pages} {1244} (\bibinfo {year}
  {1984})}\BibitemShut {NoStop}%
\bibitem [{\citenamefont {Sellitto}\ \emph {et~al.}(2005)\citenamefont
  {Sellitto}, \citenamefont {Biroli},\ and\ \citenamefont
  {Toninelli}}]{sellitto2005}%
  \BibitemOpen
  \bibfield  {author} {\bibinfo {author} {\bibfnamefont {M.}~\bibnamefont
  {Sellitto}}, \bibinfo {author} {\bibfnamefont {G.}~\bibnamefont {Biroli}}, \
  and\ \bibinfo {author} {\bibfnamefont {C.}~\bibnamefont {Toninelli}},\
  }\href@noop {} {\bibfield  {journal} {\bibinfo  {journal} {EPL}\ }\textbf
  {\bibinfo {volume} {69}},\ \bibinfo {pages} {496} (\bibinfo {year}
  {2005})}\BibitemShut {NoStop}%
\bibitem [{\citenamefont {Sellitto}\ \emph {et~al.}(2010)\citenamefont
  {Sellitto}, \citenamefont {De~Martino}, \citenamefont {Caccioli},\ and\
  \citenamefont {Arenzon}}]{sellitto2010dynamic}%
  \BibitemOpen
  \bibfield  {author} {\bibinfo {author} {\bibfnamefont {M.}~\bibnamefont
  {Sellitto}}, \bibinfo {author} {\bibfnamefont {D.}~\bibnamefont
  {De~Martino}}, \bibinfo {author} {\bibfnamefont {F.}~\bibnamefont
  {Caccioli}}, \ and\ \bibinfo {author} {\bibfnamefont {J.~J.}\ \bibnamefont
  {Arenzon}},\ }\href@noop {} {\bibfield  {journal} {\bibinfo  {journal} {Phys.
  Rev. Lett.}\ }\textbf {\bibinfo {volume} {105}},\ \bibinfo {pages} {265704}
  (\bibinfo {year} {2010})}\BibitemShut {NoStop}%
\bibitem [{\citenamefont {Ikeda}\ and\ \citenamefont
  {Miyazaki}(2015)}]{ikeda2015fre}%
  \BibitemOpen
  \bibfield  {author} {\bibinfo {author} {\bibfnamefont {H.}~\bibnamefont
  {Ikeda}}\ and\ \bibinfo {author} {\bibfnamefont {K.}~\bibnamefont
  {Miyazaki}},\ }\href@noop {} {\bibfield  {journal} {\bibinfo  {journal}
  {EPL}\ }\textbf {\bibinfo {volume} {112}},\ \bibinfo {pages} {16001}
  (\bibinfo {year} {2015})}\BibitemShut {NoStop}%
\bibitem [{\citenamefont {Sellitto}(2015)}]{sellitto2015}%
  \BibitemOpen
  \bibfield  {author} {\bibinfo {author} {\bibfnamefont {M.}~\bibnamefont
  {Sellitto}},\ }\href@noop {} {\bibfield  {journal} {\bibinfo  {journal}
  {Phys. Rev. Lett.}\ }\textbf {\bibinfo {volume} {115}},\ \bibinfo {pages}
  {225701} (\bibinfo {year} {2015})}\BibitemShut {NoStop}%
\bibitem [{\citenamefont {De~Candia}\ \emph {et~al.}(2016)\citenamefont
  {De~Candia}, \citenamefont {Fierro},\ and\ \citenamefont
  {Coniglio}}]{de2016}%
  \BibitemOpen
  \bibfield  {author} {\bibinfo {author} {\bibfnamefont {A.}~\bibnamefont
  {De~Candia}}, \bibinfo {author} {\bibfnamefont {A.}~\bibnamefont {Fierro}}, \
  and\ \bibinfo {author} {\bibfnamefont {A.}~\bibnamefont {Coniglio}},\
  }\href@noop {} {\bibfield  {journal} {\bibinfo  {journal} {Sci. Rep.}\
  }\textbf {\bibinfo {volume} {6}},\ \bibinfo {pages} {26481} (\bibinfo {year}
  {2016})}\BibitemShut {NoStop}%
\bibitem [{\citenamefont {Ikeda}\ \emph {et~al.}(2017)\citenamefont {Ikeda},
  \citenamefont {Miyazaki},\ and\ \citenamefont {Biroli}}]{ikeda2017}%
  \BibitemOpen
  \bibfield  {author} {\bibinfo {author} {\bibfnamefont {H.}~\bibnamefont
  {Ikeda}}, \bibinfo {author} {\bibfnamefont {K.}~\bibnamefont {Miyazaki}}, \
  and\ \bibinfo {author} {\bibfnamefont {G.}~\bibnamefont {Biroli}},\
  }\href@noop {} {\bibfield  {journal} {\bibinfo  {journal} {EPL}\ }\textbf
  {\bibinfo {volume} {116}},\ \bibinfo {pages} {56004} (\bibinfo {year}
  {2017})}\BibitemShut {NoStop}%
\bibitem [{\citenamefont {Berthier}\ \emph {et~al.}(2012)\citenamefont
  {Berthier}, \citenamefont {Biroli}, \citenamefont {Coslovich}, \citenamefont
  {Kob},\ and\ \citenamefont {Toninelli}}]{berthier2012finite}%
  \BibitemOpen
  \bibfield  {author} {\bibinfo {author} {\bibfnamefont {L.}~\bibnamefont
  {Berthier}}, \bibinfo {author} {\bibfnamefont {G.}~\bibnamefont {Biroli}},
  \bibinfo {author} {\bibfnamefont {D.}~\bibnamefont {Coslovich}}, \bibinfo
  {author} {\bibfnamefont {W.}~\bibnamefont {Kob}}, \ and\ \bibinfo {author}
  {\bibfnamefont {C.}~\bibnamefont {Toninelli}},\ }\href@noop {} {\bibfield
  {journal} {\bibinfo  {journal} {Phys. Rev. E}\ }\textbf {\bibinfo {volume}
  {86}},\ \bibinfo {pages} {031502} (\bibinfo {year} {2012})}\BibitemShut
  {NoStop}%
\bibitem [{\citenamefont {Wyart}\ and\ \citenamefont
  {Cates}(2017)}]{wyart2017does}%
  \BibitemOpen
  \bibfield  {author} {\bibinfo {author} {\bibfnamefont {M.}~\bibnamefont
  {Wyart}}\ and\ \bibinfo {author} {\bibfnamefont {M.~E.}\ \bibnamefont
  {Cates}},\ }\href@noop {} {\bibfield  {journal} {\bibinfo  {journal} {Phys.
  Rev. Lett.}\ }\textbf {\bibinfo {volume} {119}},\ \bibinfo {pages} {195501}
  (\bibinfo {year} {2017})}\BibitemShut {NoStop}%
\bibitem [{\citenamefont {Gazzillo}\ and\ \citenamefont
  {Pastore}(1989)}]{gazzillo1989}%
  \BibitemOpen
  \bibfield  {author} {\bibinfo {author} {\bibfnamefont {D.}~\bibnamefont
  {Gazzillo}}\ and\ \bibinfo {author} {\bibfnamefont {G.}~\bibnamefont
  {Pastore}},\ }\href@noop {} {\bibfield  {journal} {\bibinfo  {journal} {Chem.
  Phys. Lett.}\ }\textbf {\bibinfo {volume} {159}},\ \bibinfo {pages} {388}
  (\bibinfo {year} {1989})}\BibitemShut {NoStop}%
\bibitem [{\citenamefont {Grigera}\ and\ \citenamefont
  {Parisi}(2001)}]{PhysRevE.63.045102}%
  \BibitemOpen
  \bibfield  {author} {\bibinfo {author} {\bibfnamefont {T.~S.}\ \bibnamefont
  {Grigera}}\ and\ \bibinfo {author} {\bibfnamefont {G.}~\bibnamefont
  {Parisi}},\ }\href {\doibase 10.1103/PhysRevE.63.045102} {\bibfield
  {journal} {\bibinfo  {journal} {Phys. Rev. E}\ }\textbf {\bibinfo {volume}
  {63}},\ \bibinfo {pages} {045102} (\bibinfo {year} {2001})}\BibitemShut
  {NoStop}%
\bibitem [{\citenamefont {Cavagna}\ \emph {et~al.}(2012)\citenamefont
  {Cavagna}, \citenamefont {Grigera},\ and\ \citenamefont
  {Verrocchio}}]{cavagna2012}%
  \BibitemOpen
  \bibfield  {author} {\bibinfo {author} {\bibfnamefont {A.}~\bibnamefont
  {Cavagna}}, \bibinfo {author} {\bibfnamefont {T.~S.}\ \bibnamefont
  {Grigera}}, \ and\ \bibinfo {author} {\bibfnamefont {P.}~\bibnamefont
  {Verrocchio}},\ }\href@noop {} {\bibfield  {journal} {\bibinfo  {journal} {J.
  Chem. Phys.}\ }\textbf {\bibinfo {volume} {136}},\ \bibinfo {pages} {204502}
  (\bibinfo {year} {2012})}\BibitemShut {NoStop}%
\bibitem [{\citenamefont {Guti{\'e}rrez}\ \emph {et~al.}(2015)\citenamefont
  {Guti{\'e}rrez}, \citenamefont {Karmakar}, \citenamefont {Pollack},\ and\
  \citenamefont {Procaccia}}]{gutierrez2015}%
  \BibitemOpen
  \bibfield  {author} {\bibinfo {author} {\bibfnamefont {R.}~\bibnamefont
  {Guti{\'e}rrez}}, \bibinfo {author} {\bibfnamefont {S.}~\bibnamefont
  {Karmakar}}, \bibinfo {author} {\bibfnamefont {Y.~G.}\ \bibnamefont
  {Pollack}}, \ and\ \bibinfo {author} {\bibfnamefont {I.}~\bibnamefont
  {Procaccia}},\ }\href@noop {} {\bibfield  {journal} {\bibinfo  {journal}
  {EPL}\ }\textbf {\bibinfo {volume} {111}},\ \bibinfo {pages} {56009}
  (\bibinfo {year} {2015})}\BibitemShut {NoStop}%
\bibitem [{\citenamefont {Berthier}\ \emph {et~al.}(2016)\citenamefont
  {Berthier}, \citenamefont {Coslovich}, \citenamefont {Ninarello},\ and\
  \citenamefont {Ozawa}}]{berthier2016}%
  \BibitemOpen
  \bibfield  {author} {\bibinfo {author} {\bibfnamefont {L.}~\bibnamefont
  {Berthier}}, \bibinfo {author} {\bibfnamefont {D.}~\bibnamefont {Coslovich}},
  \bibinfo {author} {\bibfnamefont {A.}~\bibnamefont {Ninarello}}, \ and\
  \bibinfo {author} {\bibfnamefont {M.}~\bibnamefont {Ozawa}},\ }\href@noop {}
  {\bibfield  {journal} {\bibinfo  {journal} {Phys. Rev. Lett.}\ }\textbf
  {\bibinfo {volume} {116}},\ \bibinfo {pages} {238002} (\bibinfo {year}
  {2016})}\BibitemShut {NoStop}%
\bibitem [{\citenamefont {Ninarello}\ \emph {et~al.}(2017)\citenamefont
  {Ninarello}, \citenamefont {Berthier},\ and\ \citenamefont
  {Coslovich}}]{PhysRevX.7.021039}%
  \BibitemOpen
  \bibfield  {author} {\bibinfo {author} {\bibfnamefont {A.}~\bibnamefont
  {Ninarello}}, \bibinfo {author} {\bibfnamefont {L.}~\bibnamefont {Berthier}},
  \ and\ \bibinfo {author} {\bibfnamefont {D.}~\bibnamefont {Coslovich}},\
  }\href {\doibase 10.1103/PhysRevX.7.021039} {\bibfield  {journal} {\bibinfo
  {journal} {Phys. Rev. X}\ }\textbf {\bibinfo {volume} {7}},\ \bibinfo {pages}
  {021039} (\bibinfo {year} {2017})}\BibitemShut {NoStop}%
\bibitem [{\citenamefont {Berthier}\ \emph {et~al.}(2017)\citenamefont
  {Berthier}, \citenamefont {Charbonneau}, \citenamefont {Coslovich},
  \citenamefont {Ninarello}, \citenamefont {Ozawa},\ and\ \citenamefont
  {Yaida}}]{berthier2017breaking}%
  \BibitemOpen
  \bibfield  {author} {\bibinfo {author} {\bibfnamefont {L.}~\bibnamefont
  {Berthier}}, \bibinfo {author} {\bibfnamefont {P.}~\bibnamefont
  {Charbonneau}}, \bibinfo {author} {\bibfnamefont {D.}~\bibnamefont
  {Coslovich}}, \bibinfo {author} {\bibfnamefont {A.}~\bibnamefont
  {Ninarello}}, \bibinfo {author} {\bibfnamefont {M.}~\bibnamefont {Ozawa}}, \
  and\ \bibinfo {author} {\bibfnamefont {S.}~\bibnamefont {Yaida}},\
  }\href@noop {} {\bibfield  {journal} {\bibinfo  {journal} {PNAS}\ ,\ \bibinfo
  {pages} {201706860}} (\bibinfo {year} {2017})}\BibitemShut {NoStop}%
\bibitem [{\citenamefont {Mari}\ and\ \citenamefont
  {Kurchan}(2011)}]{mari2011}%
  \BibitemOpen
  \bibfield  {author} {\bibinfo {author} {\bibfnamefont {R.}~\bibnamefont
  {Mari}}\ and\ \bibinfo {author} {\bibfnamefont {J.}~\bibnamefont {Kurchan}},\
  }\href@noop {} {\bibfield  {journal} {\bibinfo  {journal} {J. Chem. Phys.}\
  }\textbf {\bibinfo {volume} {135}},\ \bibinfo {pages} {124504} (\bibinfo
  {year} {2011})}\BibitemShut {NoStop}%
\bibitem [{\citenamefont {Charbonneau}\ \emph {et~al.}(2014)\citenamefont
  {Charbonneau}, \citenamefont {Jin}, \citenamefont {Parisi},\ and\
  \citenamefont {Zamponi}}]{charbonneau2014b}%
  \BibitemOpen
  \bibfield  {author} {\bibinfo {author} {\bibfnamefont {P.}~\bibnamefont
  {Charbonneau}}, \bibinfo {author} {\bibfnamefont {Y.}~\bibnamefont {Jin}},
  \bibinfo {author} {\bibfnamefont {G.}~\bibnamefont {Parisi}}, \ and\ \bibinfo
  {author} {\bibfnamefont {F.}~\bibnamefont {Zamponi}},\ }\href@noop {}
  {\bibfield  {journal} {\bibinfo  {journal} {PNAS}\ }\textbf {\bibinfo
  {volume} {111}},\ \bibinfo {pages} {15025} (\bibinfo {year}
  {2014})}\BibitemShut {NoStop}%
\bibitem [{\citenamefont {Monasson}(1995)}]{monasson1995}%
  \BibitemOpen
  \bibfield  {author} {\bibinfo {author} {\bibfnamefont {R.}~\bibnamefont
  {Monasson}},\ }\href@noop {} {\bibfield  {journal} {\bibinfo  {journal}
  {Phys. Rev. Lett.}\ }\textbf {\bibinfo {volume} {75}},\ \bibinfo {pages}
  {2847} (\bibinfo {year} {1995})}\BibitemShut {NoStop}%
\bibitem [{\citenamefont {Coluzzi}\ \emph {et~al.}(1999)\citenamefont
  {Coluzzi}, \citenamefont {M{\'e}zard}, \citenamefont {Parisi},\ and\
  \citenamefont {Verrocchio}}]{coluzzi1999}%
  \BibitemOpen
  \bibfield  {author} {\bibinfo {author} {\bibfnamefont {B.}~\bibnamefont
  {Coluzzi}}, \bibinfo {author} {\bibfnamefont {M.}~\bibnamefont {M{\'e}zard}},
  \bibinfo {author} {\bibfnamefont {G.}~\bibnamefont {Parisi}}, \ and\ \bibinfo
  {author} {\bibfnamefont {P.}~\bibnamefont {Verrocchio}},\ }\href@noop {}
  {\bibfield  {journal} {\bibinfo  {journal} {J. Chem. Phys.}\ }\textbf
  {\bibinfo {volume} {111}},\ \bibinfo {pages} {9039} (\bibinfo {year}
  {1999})}\BibitemShut {NoStop}%
\bibitem [{\citenamefont {Ikeda}\ \emph {et~al.}(2016)\citenamefont {Ikeda},
  \citenamefont {Miyazaki},\ and\ \citenamefont {Ikeda}}]{ikeda2016note}%
  \BibitemOpen
  \bibfield  {author} {\bibinfo {author} {\bibfnamefont {H.}~\bibnamefont
  {Ikeda}}, \bibinfo {author} {\bibfnamefont {K.}~\bibnamefont {Miyazaki}}, \
  and\ \bibinfo {author} {\bibfnamefont {A.}~\bibnamefont {Ikeda}},\
  }\href@noop {} {\bibfield  {journal} {\bibinfo  {journal} {J. Chem. Phys.}\
  }\textbf {\bibinfo {volume} {145}},\ \bibinfo {pages} {216101} (\bibinfo
  {year} {2016})}\BibitemShut {NoStop}%
\bibitem [{\citenamefont {M\'ezard}\ \emph {et~al.}(1987)\citenamefont
  {M\'ezard}, \citenamefont {Parisi},\ and\ \citenamefont
  {Virasoo}}]{megard1987}%
  \BibitemOpen
  \bibfield  {author} {\bibinfo {author} {\bibfnamefont {M.}~\bibnamefont
  {M\'ezard}}, \bibinfo {author} {\bibfnamefont {G.}~\bibnamefont {Parisi}}, \
  and\ \bibinfo {author} {\bibfnamefont {M.~A.}\ \bibnamefont {Virasoo}},\
  }\href@noop {} {\emph {\bibinfo {title} {Spin glass theory and beyond}}}\
  (\bibinfo  {publisher} {World Scientific, Singapore},\ \bibinfo {year}
  {1987})\BibitemShut {NoStop}%
\bibitem [{\citenamefont {Nishimori}(2001)}]{nishimori2001}%
  \BibitemOpen
  \bibfield  {author} {\bibinfo {author} {\bibfnamefont {H.}~\bibnamefont
  {Nishimori}},\ }\href@noop {} {\emph {\bibinfo {title} {Statistical physics
  of spin glasses and information processing: an introduction}}},\ Vol.\
  \bibinfo {volume} {111}\ (\bibinfo  {publisher} {Clarendon Press},\ \bibinfo
  {year} {2001})\BibitemShut {NoStop}%
\bibitem [{\citenamefont {{Hajime Yoshino}}()}]{yoshino}%
  \BibitemOpen
  \bibfield  {author} {\bibinfo {author} {\bibnamefont {{Hajime Yoshino}}},\
  }\href@noop {} {\enquote {\bibinfo {title} {{Statistical mechanics of glasses
  and jamming systems: the replica method and its applications}},}\ }\bibinfo
  {howpublished}
  {\url{http://www.cp.cmc.osaka-u.ac.jp/~yoshino/articles-eng.html}}\BibitemShut
  {NoStop}%
\bibitem [{\citenamefont {Hansen}\ and\ \citenamefont
  {McDonald}(1990)}]{hansen1990}%
  \BibitemOpen
  \bibfield  {author} {\bibinfo {author} {\bibfnamefont {J.-P.}\ \bibnamefont
  {Hansen}}\ and\ \bibinfo {author} {\bibfnamefont {I.~R.}\ \bibnamefont
  {McDonald}},\ }\href@noop {} {\emph {\bibinfo {title} {Theory of simple
  liquids}}}\ (\bibinfo  {publisher} {Elsevier},\ \bibinfo {year}
  {1990})\BibitemShut {NoStop}%
\bibitem [{\citenamefont {Scala}\ \emph {et~al.}(2000)\citenamefont {Scala},
  \citenamefont {Starr}, \citenamefont {La~Nave}, \citenamefont {Sciortino},\
  and\ \citenamefont {Stanley}}]{scala1999}%
  \BibitemOpen
  \bibfield  {author} {\bibinfo {author} {\bibfnamefont {A.}~\bibnamefont
  {Scala}}, \bibinfo {author} {\bibfnamefont {F.~W.}\ \bibnamefont {Starr}},
  \bibinfo {author} {\bibfnamefont {E.}~\bibnamefont {La~Nave}}, \bibinfo
  {author} {\bibfnamefont {F.}~\bibnamefont {Sciortino}}, \ and\ \bibinfo
  {author} {\bibfnamefont {H.~E.}\ \bibnamefont {Stanley}},\ }\href@noop {}
  {\bibfield  {journal} {\bibinfo  {journal} {Nature}\ }\textbf {\bibinfo
  {volume} {406}},\ \bibinfo {pages} {166} (\bibinfo {year}
  {2000})}\BibitemShut {NoStop}%
\bibitem [{\citenamefont {Sciortino}\ \emph {et~al.}(1999)\citenamefont
  {Sciortino}, \citenamefont {Kob},\ and\ \citenamefont
  {Tartaglia}}]{PhysRevLett.83.3214}%
  \BibitemOpen
  \bibfield  {author} {\bibinfo {author} {\bibfnamefont {F.}~\bibnamefont
  {Sciortino}}, \bibinfo {author} {\bibfnamefont {W.}~\bibnamefont {Kob}}, \
  and\ \bibinfo {author} {\bibfnamefont {P.}~\bibnamefont {Tartaglia}},\ }\href
  {\doibase 10.1103/PhysRevLett.83.3214} {\bibfield  {journal} {\bibinfo
  {journal} {Phys. Rev. Lett.}\ }\textbf {\bibinfo {volume} {83}},\ \bibinfo
  {pages} {3214} (\bibinfo {year} {1999})}\BibitemShut {NoStop}%
\bibitem [{\citenamefont {Angelani}\ and\ \citenamefont
  {Foffi}(2007)}]{angelani2007}%
  \BibitemOpen
  \bibfield  {author} {\bibinfo {author} {\bibfnamefont {L.}~\bibnamefont
  {Angelani}}\ and\ \bibinfo {author} {\bibfnamefont {G.}~\bibnamefont
  {Foffi}},\ }\href@noop {} {\bibfield  {journal} {\bibinfo  {journal} {J.
  Phys. Condens. Matter}\ }\textbf {\bibinfo {volume} {19}},\ \bibinfo {pages}
  {256207} (\bibinfo {year} {2007})}\BibitemShut {NoStop}%
\bibitem [{\citenamefont {Ozawa}\ and\ \citenamefont
  {Berthier}(2017)}]{ozawa2017}%
  \BibitemOpen
  \bibfield  {author} {\bibinfo {author} {\bibfnamefont {M.}~\bibnamefont
  {Ozawa}}\ and\ \bibinfo {author} {\bibfnamefont {L.}~\bibnamefont
  {Berthier}},\ }\href@noop {} {\bibfield  {journal} {\bibinfo  {journal} {J.
  Chem. Phys.}\ }\textbf {\bibinfo {volume} {146}},\ \bibinfo {pages} {014502}
  (\bibinfo {year} {2017})}\BibitemShut {NoStop}%
\bibitem [{\citenamefont {Franz}\ and\ \citenamefont
  {Parisi}(1997)}]{PhysRevLett.79.2486}%
  \BibitemOpen
  \bibfield  {author} {\bibinfo {author} {\bibfnamefont {S.}~\bibnamefont
  {Franz}}\ and\ \bibinfo {author} {\bibfnamefont {G.}~\bibnamefont {Parisi}},\
  }\href {\doibase 10.1103/PhysRevLett.79.2486} {\bibfield  {journal} {\bibinfo
   {journal} {Phys. Rev. Lett.}\ }\textbf {\bibinfo {volume} {79}},\ \bibinfo
  {pages} {2486} (\bibinfo {year} {1997})}\BibitemShut {NoStop}%
\bibitem [{\citenamefont {Berthier}\ and\ \citenamefont
  {Coslovich}(2014)}]{berthier2014novel}%
  \BibitemOpen
  \bibfield  {author} {\bibinfo {author} {\bibfnamefont {L.}~\bibnamefont
  {Berthier}}\ and\ \bibinfo {author} {\bibfnamefont {D.}~\bibnamefont
  {Coslovich}},\ }\href@noop {} {\bibfield  {journal} {\bibinfo  {journal}
  {PNAS}\ }\textbf {\bibinfo {volume} {111}},\ \bibinfo {pages} {11668}
  (\bibinfo {year} {2014})}\BibitemShut {NoStop}%
\bibitem [{\citenamefont {Biroli}\ and\ \citenamefont
  {Bouchaud}(2012)}]{biroli2012random}%
  \BibitemOpen
  \bibfield  {author} {\bibinfo {author} {\bibfnamefont {G.}~\bibnamefont
  {Biroli}}\ and\ \bibinfo {author} {\bibfnamefont {J.-P.}\ \bibnamefont
  {Bouchaud}},\ }in\ \href@noop {} {\emph {\bibinfo {booktitle} {Structural
  Glasses and Supercooled Liquids: Theory, Experiment, and Applications}}}\
  (\bibinfo  {publisher} {John Wiley \& Sons},\ \bibinfo {year} {2012})\ pp.\
  \bibinfo {pages} {31--113}\BibitemShut {NoStop}%
\bibitem [{\citenamefont {Berthier}\ \emph {et~al.}(2004)\citenamefont
  {Berthier}, \citenamefont {Chandler},\ and\ \citenamefont
  {Garrahan}}]{berthier2004length}%
  \BibitemOpen
  \bibfield  {author} {\bibinfo {author} {\bibfnamefont {L.}~\bibnamefont
  {Berthier}}, \bibinfo {author} {\bibfnamefont {D.}~\bibnamefont {Chandler}},
  \ and\ \bibinfo {author} {\bibfnamefont {J.~P.}\ \bibnamefont {Garrahan}},\
  }\href@noop {} {\bibfield  {journal} {\bibinfo  {journal} {EPL}\ }\textbf
  {\bibinfo {volume} {69}},\ \bibinfo {pages} {320} (\bibinfo {year}
  {2004})}\BibitemShut {NoStop}%
\bibitem [{\citenamefont {Candelier}\ \emph {et~al.}(2010)\citenamefont
  {Candelier}, \citenamefont {Widmer-Cooper}, \citenamefont {Kummerfeld},
  \citenamefont {Dauchot}, \citenamefont {Biroli}, \citenamefont {Harrowell},\
  and\ \citenamefont {Reichman}}]{PhysRevLett.105.135702}%
  \BibitemOpen
  \bibfield  {author} {\bibinfo {author} {\bibfnamefont {R.}~\bibnamefont
  {Candelier}}, \bibinfo {author} {\bibfnamefont {A.}~\bibnamefont
  {Widmer-Cooper}}, \bibinfo {author} {\bibfnamefont {J.~K.}\ \bibnamefont
  {Kummerfeld}}, \bibinfo {author} {\bibfnamefont {O.}~\bibnamefont {Dauchot}},
  \bibinfo {author} {\bibfnamefont {G.}~\bibnamefont {Biroli}}, \bibinfo
  {author} {\bibfnamefont {P.}~\bibnamefont {Harrowell}}, \ and\ \bibinfo
  {author} {\bibfnamefont {D.~R.}\ \bibnamefont {Reichman}},\ }\href {\doibase
  10.1103/PhysRevLett.105.135702} {\bibfield  {journal} {\bibinfo  {journal}
  {Phys. Rev. Lett.}\ }\textbf {\bibinfo {volume} {105}},\ \bibinfo {pages}
  {135702} (\bibinfo {year} {2010})}\BibitemShut {NoStop}%
\bibitem [{\citenamefont {Ciamarra}\ \emph {et~al.}(2016)\citenamefont
  {Ciamarra}, \citenamefont {Pastore},\ and\ \citenamefont
  {Coniglio}}]{ciamarra2016}%
  \BibitemOpen
  \bibfield  {author} {\bibinfo {author} {\bibfnamefont {M.~P.}\ \bibnamefont
  {Ciamarra}}, \bibinfo {author} {\bibfnamefont {R.}~\bibnamefont {Pastore}}, \
  and\ \bibinfo {author} {\bibfnamefont {A.}~\bibnamefont {Coniglio}},\
  }\href@noop {} {\bibfield  {journal} {\bibinfo  {journal} {Soft matter}\
  }\textbf {\bibinfo {volume} {12}},\ \bibinfo {pages} {358} (\bibinfo {year}
  {2016})}\BibitemShut {NoStop}%
\bibitem [{\citenamefont {Gleim}\ \emph {et~al.}(1998)\citenamefont {Gleim},
  \citenamefont {Kob},\ and\ \citenamefont {Binder}}]{PhysRevLett.81.4404}%
  \BibitemOpen
  \bibfield  {author} {\bibinfo {author} {\bibfnamefont {T.}~\bibnamefont
  {Gleim}}, \bibinfo {author} {\bibfnamefont {W.}~\bibnamefont {Kob}}, \ and\
  \bibinfo {author} {\bibfnamefont {K.}~\bibnamefont {Binder}},\ }\href
  {\doibase 10.1103/PhysRevLett.81.4404} {\bibfield  {journal} {\bibinfo
  {journal} {Phys. Rev. Lett.}\ }\textbf {\bibinfo {volume} {81}},\ \bibinfo
  {pages} {4404} (\bibinfo {year} {1998})}\BibitemShut {NoStop}%
\bibitem [{\citenamefont {Szamel}\ and\ \citenamefont
  {Flenner}(2004)}]{szamel2004}%
  \BibitemOpen
  \bibfield  {author} {\bibinfo {author} {\bibfnamefont {G.}~\bibnamefont
  {Szamel}}\ and\ \bibinfo {author} {\bibfnamefont {E.}~\bibnamefont
  {Flenner}},\ }\href@noop {} {\bibfield  {journal} {\bibinfo  {journal} {EPL}\
  }\textbf {\bibinfo {volume} {67}},\ \bibinfo {pages} {779} (\bibinfo {year}
  {2004})}\BibitemShut {NoStop}%
\bibitem [{\citenamefont {Berthier}\ and\ \citenamefont
  {Kob}(2007)}]{berthier2007mon}%
  \BibitemOpen
  \bibfield  {author} {\bibinfo {author} {\bibfnamefont {L.}~\bibnamefont
  {Berthier}}\ and\ \bibinfo {author} {\bibfnamefont {W.}~\bibnamefont {Kob}},\
  }\href@noop {} {\bibfield  {journal} {\bibinfo  {journal} {J. Phys. Condens.
  Matter}\ }\textbf {\bibinfo {volume} {19}},\ \bibinfo {pages} {205130}
  (\bibinfo {year} {2007})}\BibitemShut {NoStop}%
\bibitem [{\citenamefont {Franz}\ \emph {et~al.}(2012)\citenamefont {Franz},
  \citenamefont {Jacquin}, \citenamefont {Parisi}, \citenamefont {Urbani},\
  and\ \citenamefont {Zamponi}}]{franz2012quantitative}%
  \BibitemOpen
  \bibfield  {author} {\bibinfo {author} {\bibfnamefont {S.}~\bibnamefont
  {Franz}}, \bibinfo {author} {\bibfnamefont {H.}~\bibnamefont {Jacquin}},
  \bibinfo {author} {\bibfnamefont {G.}~\bibnamefont {Parisi}}, \bibinfo
  {author} {\bibfnamefont {P.}~\bibnamefont {Urbani}}, \ and\ \bibinfo {author}
  {\bibfnamefont {F.}~\bibnamefont {Zamponi}},\ }\href@noop {} {\bibfield
  {journal} {\bibinfo  {journal} {PNAS}\ }\textbf {\bibinfo {volume} {109}},\
  \bibinfo {pages} {18725} (\bibinfo {year} {2012})}\BibitemShut {NoStop}%
\bibitem [{\citenamefont {Ikeda}\ and\ \citenamefont
  {Miyazaki}(2011)}]{ikeda2011glass}%
  \BibitemOpen
  \bibfield  {author} {\bibinfo {author} {\bibfnamefont {A.}~\bibnamefont
  {Ikeda}}\ and\ \bibinfo {author} {\bibfnamefont {K.}~\bibnamefont
  {Miyazaki}},\ }\href@noop {} {\bibfield  {journal} {\bibinfo  {journal}
  {Phys. Rev. Lett.}\ }\textbf {\bibinfo {volume} {106}},\ \bibinfo {pages}
  {015701} (\bibinfo {year} {2011})}\BibitemShut {NoStop}%
\bibitem [{\citenamefont {Foini}\ \emph {et~al.}(2012)\citenamefont {Foini},
  \citenamefont {Krzakala},\ and\ \citenamefont {Zamponi}}]{foini2012}%
  \BibitemOpen
  \bibfield  {author} {\bibinfo {author} {\bibfnamefont {L.}~\bibnamefont
  {Foini}}, \bibinfo {author} {\bibfnamefont {F.}~\bibnamefont {Krzakala}}, \
  and\ \bibinfo {author} {\bibfnamefont {F.}~\bibnamefont {Zamponi}},\
  }\href@noop {} {\bibfield  {journal} {\bibinfo  {journal} {J. Stat. Mech.
  Theor. Exp.}\ }\textbf {\bibinfo {volume} {2012}},\ \bibinfo {pages} {P06013}
  (\bibinfo {year} {2012})}\BibitemShut {NoStop}%
\end{thebibliography}%

\end{document}